\documentclass[aps,prb,amsmath,superscriptaddress,preprint,bibnotes,longbibliography,floatfix]{revtex4-2}

\usepackage{mathptmx,newtxtext,newtxmath,xspace}
\usepackage{amsbsy,bm,bbold}
\usepackage{siunitx}
\usepackage{graphicx,color,xcolor}
\usepackage{blindtext}
\usepackage{dcolumn}
\usepackage{tabularx}
\usepackage{lineno}
\usepackage{lipsum}
\usepackage[colorlinks=true, 
            linkcolor=blue, 
            urlcolor=blue,
            citecolor=blue]{hyperref}

\newcommand{\iu}{{i\mkern1mu}}

\newcommand{\CoP}{Na$_{2}$BaCo(PO$_4$)$_{2}$}

\definecolor{darkblue}{rgb}{0.,0.,0.4}
\definecolor{darkred}{rgb}{0.5,0.,0.}
\definecolor{BlueViolet}{RGB}{138,43,226}
\definecolor{SkyBlue}{RGB}{30,144,255}
\definecolor{DarkGreen}{RGB}{0,100,0}

\begin{document}
\title{Continuum of spin excitations in an ordered magnet}

\author{Jieming~Sheng}
\thanks{These authors contributed equally to this work}
\affiliation{Department of Physics, Southern University of Science and Technology, Shenzhen 518055, China}
\affiliation{School of Physical Sciences, Great Bay University, Dongguan 523000, China}
\affiliation{Great Bay Institute for Advanced Study, Dongguan 523000, China}

\author{Le~Wang}
\thanks{These authors contributed equally to this work}
\affiliation{Department of Physics, Southern University of Science and Technology, Shenzhen 518055, China}
\affiliation{Shenzhen Institute for Quantum Science and Engineering, Shenzhen 518055, China}

\author{Wenrui~Jiang}
\affiliation{Department of Physics, Southern University of Science and Technology, Shenzhen 518055, China}
\affiliation{Shenzhen Institute for Quantum Science and Engineering, Shenzhen 518055, China}

\author{Han~Ge}
\affiliation{Department of Physics, Southern University of Science and Technology, Shenzhen 518055, China}

\author{Nan~Zhao}
\affiliation{Department of Physics, Southern University of Science and Technology, Shenzhen 518055, China}

\author{Tiantian~Li}
\affiliation{Department of Physics, Southern University of Science and Technology, Shenzhen 518055, China}

\author{Maiko~Kofu}
\affiliation{J-PARC Center, Japan Atomic Energy Agency, Tokai, Ibaraki 319-1195, Japan}

\author{Dehong~Yu}
\affiliation{Australian Nuclear Science and Technology Organisation, Lucas Heights, New South Wales 2234, Australia}

\author{Wei~Zhu}
\affiliation{School of Science, Westlake University, Hangzhou 310030, China}
\affiliation{Institute of Natural Sciences, Westlake Institute of Advanced Study, Hangzhou 310024, China}

\author{Jia-Wei~Mei}
\email[Corresponding author: ]{meijw@sustech.edu.cn}
\affiliation{Department of Physics, Southern University of Science and Technology, Shenzhen 518055, China}
\affiliation{Shenzhen Institute for Quantum Science and Engineering, Shenzhen 518055, China}
\affiliation{Shenzhen Key Laboratory of Advanced Quantum Functional Materials and Devices, Southern University of Science and Technology, Shenzhen 518055, China}

\author{Zhentao~Wang}
\email[Corresponding author: ]{ztwang@zju.edu.cn}
\affiliation{Center for Correlated Matter and School of Physics, Zhejiang University, Hangzhou 310058, China}

\author{Liusuo~Wu}
\email[Corresponding author: ]{wuls@sustech.edu.cn}
\affiliation{Department of Physics, Southern University of Science and Technology, Shenzhen 518055, China}
\affiliation{Shenzhen Key Laboratory of Advanced Quantum Functional Materials and Devices, Southern University of Science and Technology, Shenzhen 518055, China}

\date{\today}

\maketitle

\textbf{Continuum of spin excitations observed in inelastic neutron scattering experiments are often considered as a strong evidence of quantum spin liquid formation. 
When quantum spin liquid is indeed the ground state of a disorder-free magnetic compound, the elementary excitation is no longer the conventional spin waves (magnons). 
Instead, the magnons fractionalize into spinons, leaving only a two-spinon continuum detectable.
For a clean ordered antiferromagnet, it was unclear if we can observe a continuous spectrum similar to the ones in a quantum spin liquid. 
Here we show that the magnetically ordered state in \CoP{} hosts a spin excitation continuum induced by strong quantum fluctuations.
Thus, a second thought is necessary when concluding such continuum as signature of quantum spin liquid in new material explorations.
}

The quantum spin liquid (QSL) state was originally proposed by Philip W. Anderson as the ground state of the spin-1/2 triangular lattice (TL) Heisenberg model, where the combination of geometric frustration and quantum fluctuation destroys the long-range magnetic order even at zero temperature~\cite{AndersonPW1973}. 
The QSL, if realized, could provide a robust platform for fault-tolerant quantum computation~\cite{KitaevAY2003,KitaevA2006}, and may also greatly boost our understanding of the formation mechanism of unconventional superconductivity~\cite{AndersonPW1987,LeePA2006_RMP}. 
Over the years, tremendous effort has been spent in the quest for the realization of QSL in real materials~\cite{BalentsL2010_review,ZhouY2017_RMP,WenJ2019_review,BroholmC2020_review}.


While ``there is no single experimental feature that identifies a spin-liquid state''~\cite{BalentsL2010_review}, several series of materials have demonstrated potential as QSL candidates, including the organic compounds ($\kappa$-(BEDT-TTF)$_2$Cu$_2$(CN)$_3$~\cite{ShimizuY2003}, EtMe$_3$Sb[Pd(dmit)$_2$]$_2$~\cite{ItouT2008}, $\kappa$-H$_3$(Cat-EDT-TTF)$_2$~\cite{IsonoT2014}), and inorganic compounds on kagome (ZnCu$_3$(OH)$_6$Cl$_2$~\cite{HanTH2012}), triangular (YbZn$_2$GaO$_5$~\cite{XuS2023}, rare-earth chalcogenides NaYb{\it X}$_2$ where {\it X}=\{O, S, Se\}~\cite{LiuW2018,BaenitzM2018,BordelonMM2019,DaiPL2021,ScheieAO2024}, 1T-TaS$_2$~\cite{LawKT2017}), honeycomb ($\alpha$-RuCl$_3$~\cite{WangZ2017_RuCl3,ZhengJ2017}, BaCo$_2$(AsO$_4$)$_2$~\cite{HalloranT2023}), distorted bilayer kagome (Ca$_{10}$Cr$_7$O$_{28}$~\cite{BalzC2016}), hyper-kagome (Na$_4$Ir$_3$O$_8$~\cite{OkamotoY2007}) and trillium lattices (K$_2$Ni$_2$(SO$_4$)$_3$~\cite{ZivkovicI2021}).


In QSLs, the conventional spin wave excitations fractionalize into the spinon continuum~\cite{LeePA2006_RMP,ZhouY2017_RMP,ZhuW2019}, which can be directly observed in inelastic neutron scattering (INS) experiments. 
Indeed, several QSL candidates have demonstrated the continuous excitation spectrum~\cite{HanTH2012,BalzC2016,BordelonMM2019,DaiPL2021,ScheieAO2024}, although the exact origins of these continua are still under debate.
For example, the effects of the disorder may possibly play some role in the continuum formation in ZnCu$_3$(OH)$_6$Cl$_2$ and NaYbX$_2$ (X=O, S, Se).

Thus, it is worth asking whether we can directly associate the observed INS continuum with the QSL behavior for materials that do not order down to the lowest available temperature. For compounds with other active degrees of freedom (e.g. charge) or with disorder, it is hard to make definite conclusions, since the magnon lifetime can also be significantly shortened by scattering with electrons or disorder. Nevertheless, we can still ask, for good insulators without intrinsic disorder, are the fully continuous excitations always the signature of the QSL state?

In this article, we find the recently discovered spin-1/2 TL antiferromagnet \CoP{}~\cite{ZhongR2019} a great material that clarifies this long-standing question. Originally, \CoP{} was also considered a QSL candidate with thermodynamic measurements down to \qty{50}{mK} at zero field~\cite{ZhongR2019,LeeS2021}, and later a transition temperature $T_\text{N}\sim \qty{150}{mK}$ was discovered at which point the compound orders into an antiferromagnetic (AFM) state~\cite{LiN2020,ShengJ2022_NaBaCoPO}.
As we shall demonstrate later, while \CoP{} exhibits clear long-range magnetic order at low-$T$, a spin excitation continuum is observed. Through extensive experimental and theoretical analysis, we show that the continuum is not caused by the disorder effect, but actually an intrinsic behavior of the spin-1/2 XXZ model on TL. In other words, the spin excitation continuum and the long-range magnetic order are able to co-exist at the interplay of geometric frustration and quantum fluctuations.


The Hamiltonian of \CoP{} was established by comprehensive thermodynamic and spectroscopic measurements in Ref.~\cite{ShengJ2022_NaBaCoPO}, that leads to a spin-1/2 XXZ model on TL:
\begin{equation}
\mathcal{H} = J\sum_{\langle ij \rangle} \left( S_i^x S_j^x + S_i^y S_j^y + \Delta S_i^z S_j^z \right),
\label{eq:xxz}
\end{equation}
where $J=\qty{0.076(1)}{meV}$ is the nearest neighbor interaction of the TL, and the exchange anisotropy $\Delta = 1.645(1)$. The inter-layer interactions are at least an order of magnitude smaller compared to $J$, justifying the two-dimensional TL as the relevant model for most of the analysis in this paper (see Fig.~S2 in Supplementary Materials). 
Other symmetry-allowed exchange anisotropies were shown to be negligible compared to $J$~\cite{ShengJ2022_NaBaCoPO}.
Independent model extraction by the tensor network~\cite{GaoY2022_NaBaCoPO} has also obtained parameters that are very close to Ref.~\cite{ShengJ2022_NaBaCoPO}.

\begin{figure*}[t!]
\includegraphics[width=0.99\textwidth]{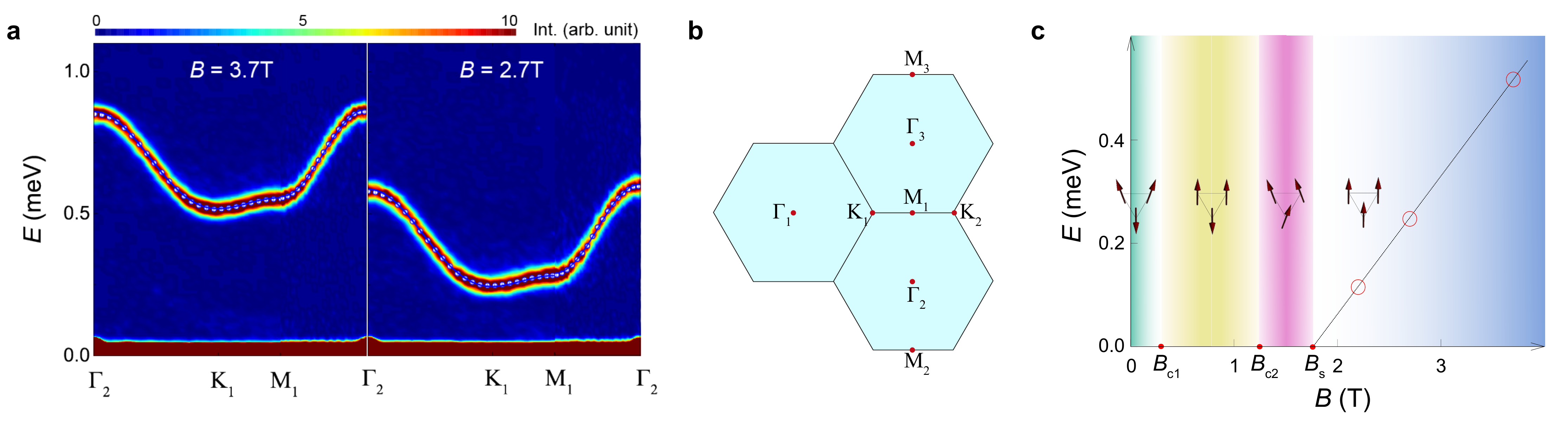}
\caption{Single-magnon BEC in the fully polarized phase. 
{\bf a} Inelastic neutron scattering results of \CoP{} for $B=\qty{3.7}{T}$ and $B=\qty{2.7}{T}$ at \qty{60}{mK}. The black empty circles are centers of the spin wave fitted by Gaussian function. The black solid line is the theoretical 1-magnon dispersion \eqref{eq:disp_FP} with $J=\qty{0.076}{meV}$, $\Delta=1.645$, and $g_c=4.645$.
{\bf b} High-symmetry momentum points used in the trajectory of {\bf a}, where $\Gamma_1=(0,0,0)$ is the center of the first Brillouin zone. {\bf c} The red empty circles are the measured spin wave gap at K point. The black solid line is the theoretical value given by Eq.~\eqref{eq:disp_FP}. The $T=0$ phases are schematically indicated along with the critical magnetic fields \{$B_\text{c1}$, $B_\text{c2}$, $B_\text{s}$\}. The softening of the spin wave gap at $B_\text{s}$ is a signature of magnon BEC.}
\label{fig:FP_SW}
\end{figure*}

We start by showing that the magnetic exchanges in \CoP{} are indeed clean nearest-neighbor XXZ type. This is achieved by fully polarizing the magnetic moments along the $c$-axis with a strong magnetic field $B>B_\text{s}$, where the linear spin wave theory becomes an exact solution that facilitates accurate model extraction. As shown in Fig.~\ref{fig:FP_SW}{\bf a}-{\bf b}, the sharp spin waves agree perfectly with the theoretical dispersion: 
\begin{equation}
\omega_{\bm{k}}^\text{(FP)}= 2JS \left( \cos k_x + 2 \cos \frac{k_x}{2} \cos \frac{\sqrt{3}k_y}{2} \right) - 6S\Delta J + g_c \mu_B B,
\label{eq:disp_FP}
\end{equation}
where $g_c$ is the g-factor along the $c$-axis, $\mu_B$ is the Bohr magneton, and the magnetic moment is $S=1/2$. Clearly, neither structural nor magnetic disorders are allowed to play a significant role in this compound; otherwise, the spin waves above $B_\text{s}$ would be broadened due to scattering between the magnons and the disorder.

By using several different magnetic fields above $B_\text{s}$ (Fig.~\ref{fig:FP_SW}), we determined the $g$-factor to be $g_c=4.645(22)$. More interestingly, the 1-magnon gap at K-point extends linearly to zero, and condenses exactly at the quantum critical point (QCP) $B=B_\text{s}$ (Fig.~\ref{fig:FP_SW}{\bf c}). In other words, the nature of the QCP is the one-magnon BEC~\cite{ZapfV2014_RMP}. The ground state slightly below $B_\text{s}$ can be readily inferred from the one-magnon state being condensed: it is both a superfluid due to magnon condensation, and a solid due to translational symmetry breaking (softened at the K-point). In other words, the phase right below $B_\text{s}$ is a supersolid in the bosonic representation~\cite{AndreevAF1969,ChesterGV1970,LeggettAJ1970}. In the spin representation, this supersolid phase has a ``V''-shape, known from cluster mean-field and density-matrix renormalization group (DMRG) calculations~\cite{YamamotoD2014,SellmannD2015}. 

Having established the TL XXZ model \eqref{eq:xxz} as an accurate description for \CoP{} and excluded the role of disorder, we are ready to explore the effects of quantum fluctuation that becomes significant at lower field. Thanks to the small saturation field in \CoP{} ($B_\text{s}=\qty{1.8}{T}$ with $B \parallel c$), our comprehensive INS measurements were able to explore all the different magnetic ground states of this easy-axis TL XXZ model with a field applied along the $c$-axis, namely the ``Y'', up-up-down (UUD), ``V'', and the fully polarized (FP) phases (see Fig.~\ref{fig:FP_SW}{\bf c} and Refs.~\cite{YamamotoD2014,SellmannD2015}).

\begin{figure*}[t!]
\includegraphics[width=0.99\textwidth]{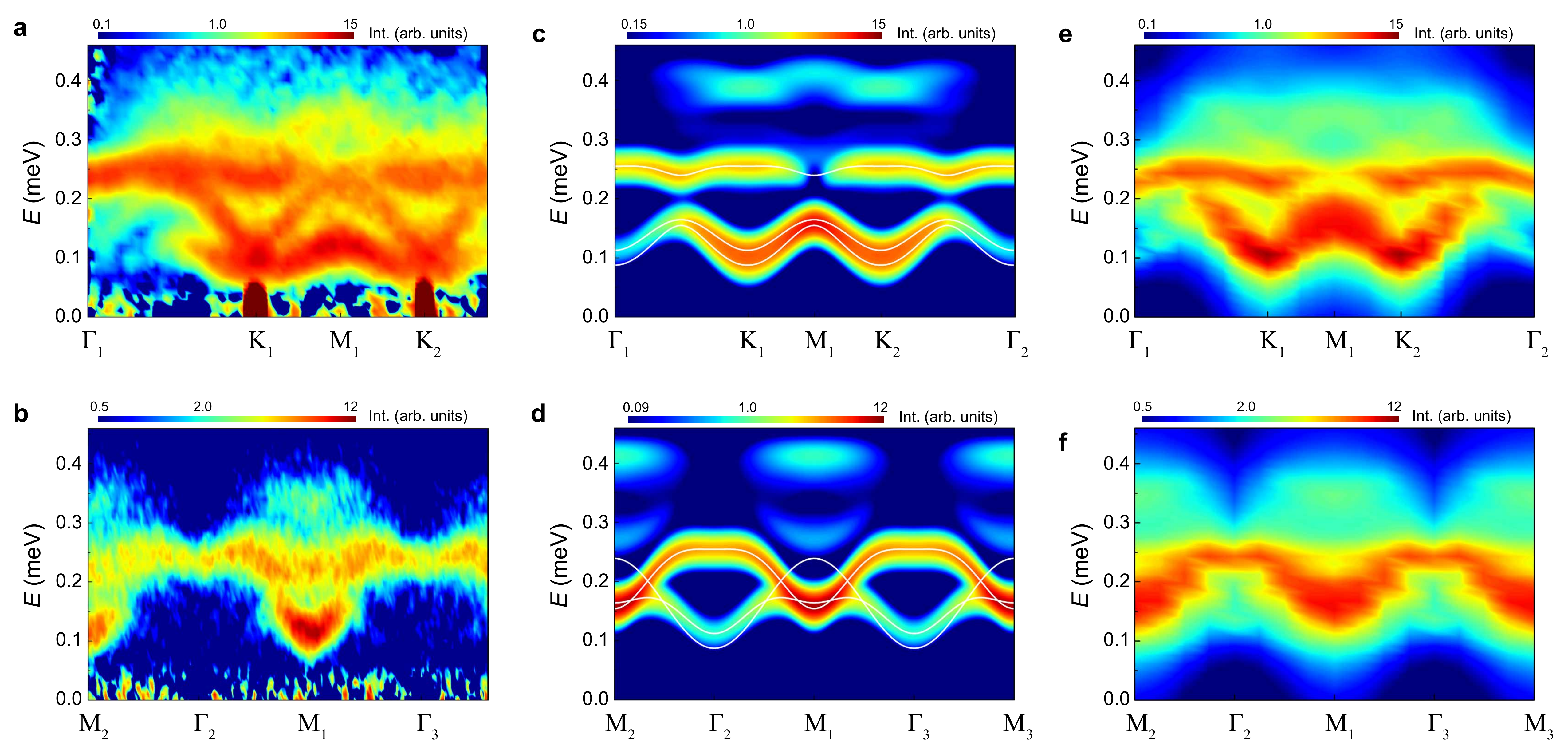}
\caption{Spin excitation spectra along the high-symmetry momentum directions at $B=\qty{0.75}{T}$ (UUD phase). {\bf a}-{\bf b} Inelastic neutron scattering results at $T=\qty{60}{mK}$ with a $\qty{60}{mK}$-$\qty{3.7}{T}$ data set subtracted as background. The incident neutron energy is $E_i=\qty{2.63}{meV}$. {\bf c}-{\bf d} $T=0$ dynamic spin structure factor $\mathcal{S}(\bm{k},\omega)$ from linear spin wave, including both the 1-magnon and 2-magnon contributions. The white solid lines are the 1-magnon dispersion. {\bf e}-{\bf f} $T=0$ dynamic spin structure factor $\mathcal{S}(\bm{k},\omega)$ from density matrix renormalization group on $6$-leg cylinder.}
\label{fig:B0d75T_SW}
\end{figure*}

The UUD phase is located on the 1/3-magnetization plateau for $\qty{0.3}{T}<B<\qty{1.2}{T}$. 
To reveal the magnetic excitation in this phase, we have performed INS measurements at $B=\qty{0.75}{T}$ (see Fig.~\ref{fig:B0d75T_SW}{\bf a}-{\bf b}). Linear spin wave (LSW) calculations (see methods) are shown in Fig.~\ref{fig:B0d75T_SW}{\bf c}-{\bf d
} for comparison. The main features, including three magnon branches with a global minimum at K-points, are qualitatively captured by LSW. 

Some fine features are not accurately reproduced by LSW. For instance, the measured highest spin wave branch goes up as we start leaving the $\Gamma_1$-point towards $\text{K}_1$ direction (Fig.~\ref{fig:B0d75T_SW}{\bf a}), opposite to the LSW prediction (Fig.~\ref{fig:B0d75T_SW}{\bf c}). The 2-magnon continuum from INS measurements, mainly between 0.3 and \qty{0.4}{meV}, also behaves slightly differently from the LSW calculation.  

To see if such differences are originated from an inaccurate modeling, or from higher order quantum fluctuations that modify the spin wave spectrum, we use the density matrix renormalization group (DMRG) to compute the dynamic spin structure factor~\cite{KuhnerTD1999,JeckelmannE2002}.
As shown in Fig.~\ref{fig:B0d75T_SW}{\bf e}-{\bf f}, DMRG indeed renormalizes the LSW spectrum to better represent the experimental one. The initial slope of the highest magnon branch near $\Gamma$-point is corrected. The 2-magnon continuum in the LSW results are shifted downwards, becoming almost quantitatively the same as the INS ones. The middle magnon branch near M-point is also shifted upwards that becomes consistent with the INS results, although within our resolution it is still not fully separated from the lowest branch (Fig.~\ref{fig:B0d75T_SW}{\bf e}).


The comparison of the experimental and theoretical results of the UUD phase deliver a few important messages. Firstly, the nearest-neighbor-only XXZ model~\eqref{eq:xxz} is again confirmed as an accurate microscopic model for \CoP{}. As expected from the INS results in the FP phase, the absence of disorder in our compound also leads to sharp 1-magnon excitations in the UUD phase. Secondly, the quantum fluctuations play an important role in the spin excitation spectrum, resulting in not only renormalization of the 1-magnon dispersion, but also transfers significant weight from the 1-magnon branch into the 2-magnon continuum. While the 2-magnon continuum is present, we note that it does not lead to any obvious decay of the 1-magnon branches. This is an expected behavior since the collinear UUD state does not produce any 3-magnon interaction that is responsible for the 1- to 2-magnon decays~\cite{ZhitomirskyME2013_RMP}. Lastly, we note that a standard LSW would lead to systematic error in fitting the 1-magnon bands in the UUD phase, so one should rely on more accurate fits in the FP phase or use more accurate methods (e.g., DMRG) if $B_\text{s}$ is large.

\begin{figure*}[t!]
\includegraphics[width=0.99\textwidth]{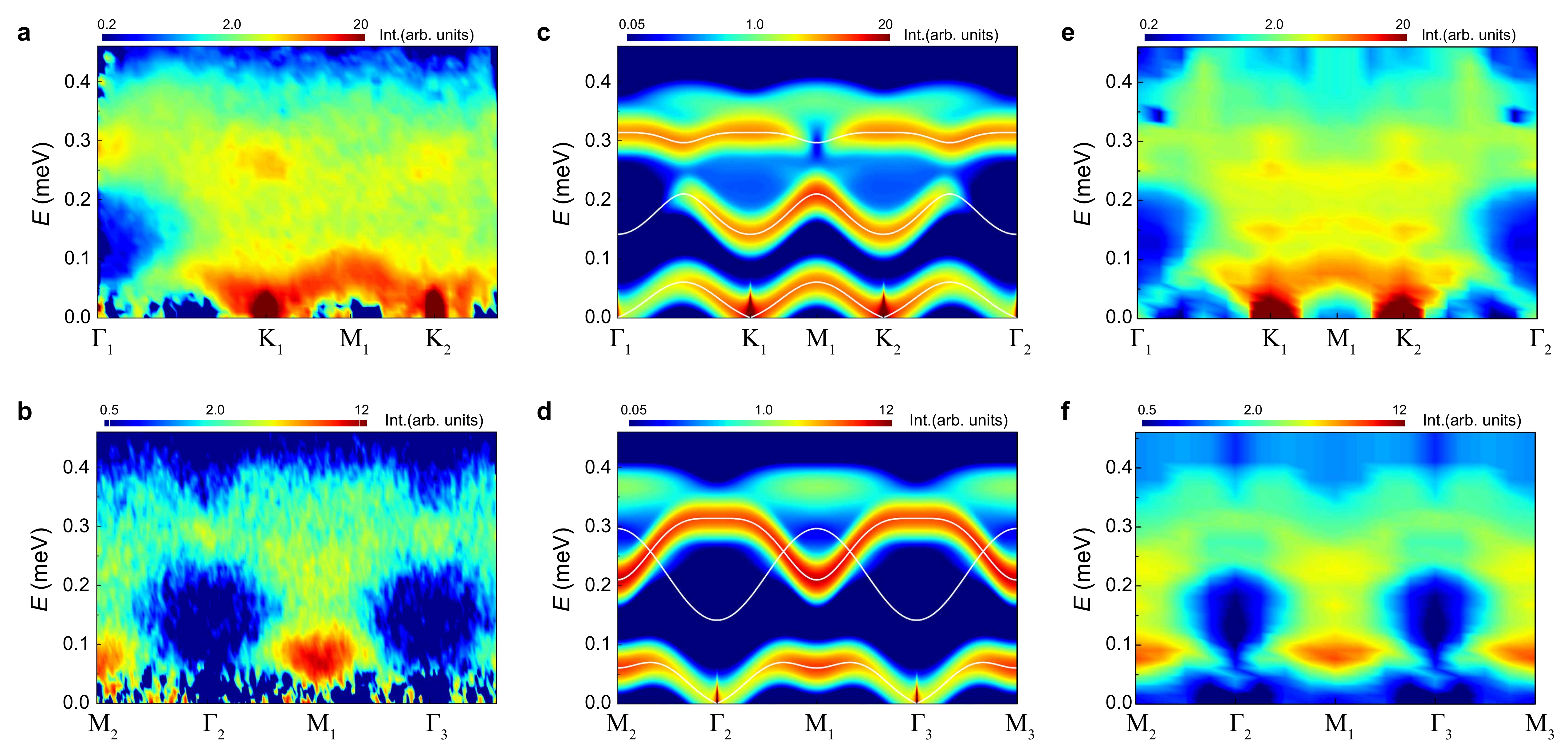}
\caption{Spin excitation spectra along the high-symmetry momentum directions at $B=\qty{1.2}{T}$ (V phase). {\bf a}-{\bf b} Inelastic neutron scattering results at $T=\qty{60}{mK}$ with a $\qty{60}{mK}$-$\qty{3.7}{T}$ data set subtracted as background. The incident neutron energy is $E_i=\qty{2.63}{meV}$. {\bf c}-{\bf d} $T=0$ dynamic spin structure factor $\mathcal{S}(\bm{k},\omega)$ from linear spin wave, including both the 1-magnon and 2-magnon contributions. The white solid lines are the 1-magnon dispersion. {\bf e}-{\bf f} $T=0$ dynamic spin structure factor $\mathcal{S}(\bm{k},\omega)$ from density matrix renormalization group on $6$-leg cylinder.}
\label{fig:B1d2T_SW}
\end{figure*}

The condition of zero 3-magnon interaction in the UUD state no longer holds in noncollinear states, including both the high-field ``V'' and low-field ``Y'' phases. 
Figure~\ref{fig:B1d2T_SW} shows both the experimental and theoretical spin excitation spectra at \qty{1.2}{T} and \qty{60}{mK}, where the ground state is the ``V'' phase.  While the Goldstone mode at K-points remains sharply defined, the higher-energy modes from INS (Fig.~\ref{fig:B1d2T_SW}{\bf a}-{\bf b}) are significantly broadened, where the decay into 2-magnon continuum is possibly responsible. Indeed, the LSW results shows significant overlap of the higher-energy modes with the 2-magnon continuum (Fig.~\ref{fig:B1d2T_SW}), allowing this decay channel. Again, the DMRG results quantitatively agree with the INS ones, confirming that such broadening is an intrinsic feature of the TL XXZ model rather than being induced by disorder.

\begin{figure*}[t!]
\includegraphics[width=0.99\textwidth]{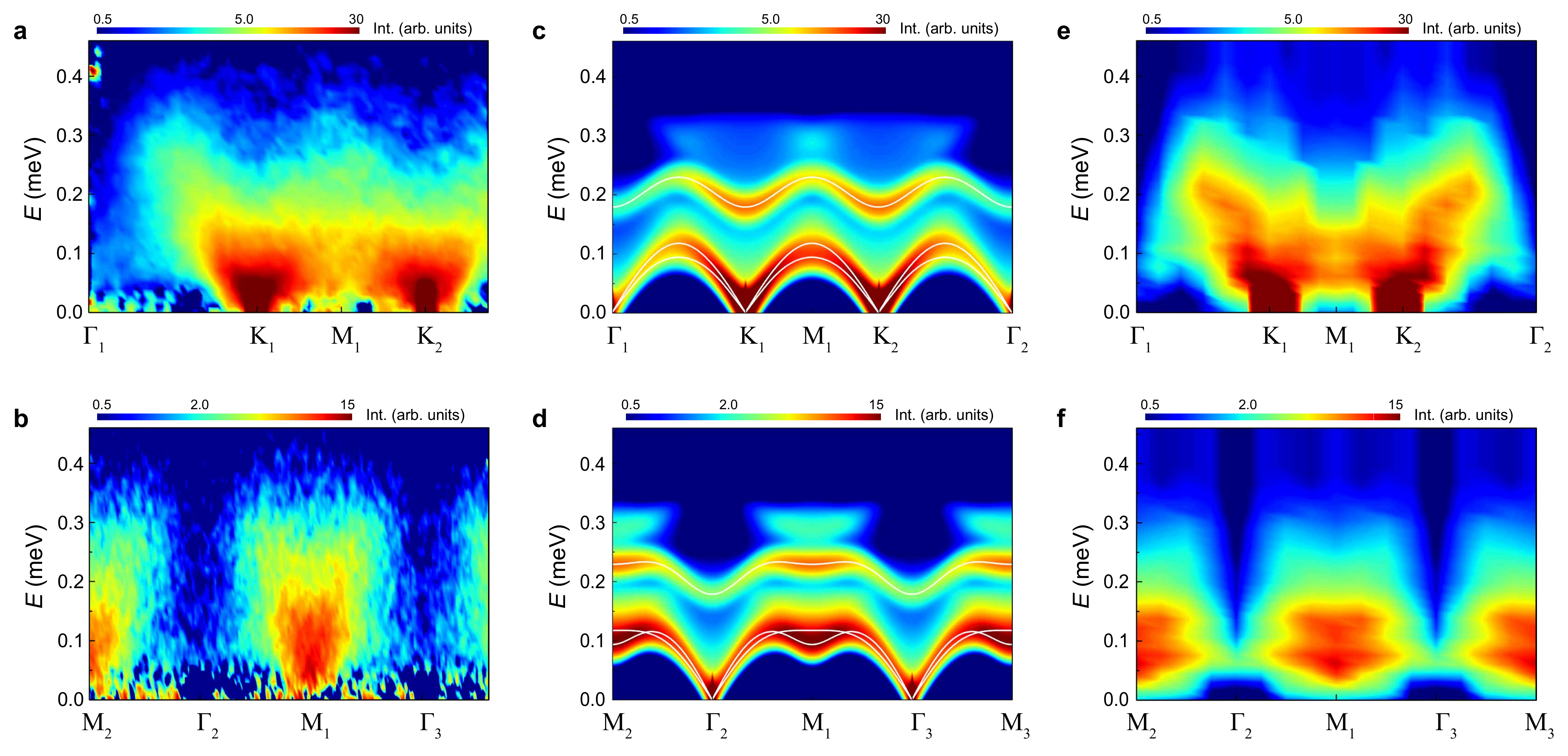}
\caption{Spin excitation spectra along the high-symmetry momentum directions at $B=\qty{0}{T}$ (Y phase). {\bf a}-{\bf b} Inelastic neutron scattering results at $T=\qty{60}{mK}$ with a $\qty{60}{mK}$-$\qty{3.7}{T}$ data set subtracted as background. The incident neutron energy is $E_i=\qty{2.63}{meV}$. {\bf c}-{\bf d} $T=0$ dynamic spin structure factor $\mathcal{S}(\bm{k},\omega)$ from linear spin wave, including both the 1-magnon and 2-magnon contributions. The white solid lines are the 1-magnon dispersion. {\bf e}-{\bf f} $T=0$ dynamic spin structure factor $\mathcal{S}(\bm{k},\omega)$ from density matrix renormalization group on $6$-leg cylinder.}
\label{fig:B0T_SW}
\end{figure*}

Finally, the zero-field INS results are shown in Fig.~\ref{fig:B0T_SW}{\bf a}-{\bf b}. The measurements were taken at $T=\qty{60}{mK}$, much lower than the N\'eel temperature $T_\text{N}=\qty{150}{mK}$ below which the long-range ``Y'' phase is stabilized. Normally, for such magnetically ordered state, at least some remaining sharp spin waves are expected. However, in contrast to our intuition, there is practically no sharply defined spin waves in Fig.~\ref{fig:B0T_SW}{\bf a}-{\bf b}, and only a continuum of spin excitations is visible. 
This continuum was first observed on PELICAN at ANSTO with incident energy $E_i=\qty{3.7}{meV}$ and \qty{0.13}{meV} energy resolution (see Supplemental Materials Fig.~S1). Later we further confirmed this continuum on AMATERAS at J-PARC with better energy resolution of  \qty{0.047}{meV} and $E_i=\qty{2.63}{meV}$.

If one had not performed the prior comprehensive analysis that both firmly established the long-range magnetic order and excluded the disorder effect, a ``standard'' practice would be concluding such continuum as a ``smoking-gun evidence'' of a QSL. Clearly, the QSL proposal is out of the question for \CoP{}, and the spin wave broadening must be an intrinsic feature of the TL XXZ model in the ordered ``Y'' phase.

In fact, the LSW shows that all the 1-magnon branches are strongly overlapping with the 2-magnon continuum (Fig.~\ref{fig:B0T_SW}{\bf c}-{\bf d}), suggesting that they are unstable due to the decay process. This is confirmed by the DMRG calculation, shown in Fig.~\ref{fig:B0T_SW}{\bf e}-{\bf f}, where the sharp spin waves decay into a continuum of spin excitations that agree with the INS measurements.


From a different perspective, the continuum formation is also a natural consequence due to proximity to a QCP beyond which point the long-range magnetic order is destroyed (in the Heisenberg limit $\Delta \rightarrow 1$, a tiny next-nearest-neighbor interaction that is about 6\% of the nearest-neighbor is able to induce a QSL phase~\cite{ZhuZ2015,HuWJ2015}). In other words, the free spinons on the QSL side now becomes bounded into magnons on the ordered side, and such magnons preserve the internal two-spinon structures if the system is not too far from the QCP. As a result, the continuum could be a feature of the rich internal spinon-pair-like structures of magnons~\cite{GhioldiEA2018}. For \CoP{}, the QSL on the other side of the putative QCP is argued to be the Dirac type~\cite{JiaH2023}.

It is interesting to note that a few other TL compounds have also demonstrated similar continuous spin excitation spectra recently, particularly including the series of AYbX$_2$~\cite{DaiPL2021,XieT2023,ScheieAO2023,ScheieAO2024}. While those continua may also arise from similar physics, we argue that \CoP{} is advantageous in revealing the intrinsic nature of quantum fluctuations that causes the continuum formation. 
We note that neither NaYbSe$_2$ nor CsYbSe$_2$ exhibits true long-range order down to the lowest available temperature; while KYbSe$_2$ shows a specific heat anomaly near \qty{290}{mK}, the neutron diffraction does not reveal magnetic Bragg peak along $L$-direction below this temperature~\cite{ScheieAO2024}. Furthermore, NaYbSe$_2$ likely suffers from site disorder.
As a result, they are not the best case to single out quantum fluctuation as the only source of the observed continua in a long-range ordered state.

To summarize, we have performed INS and theoretical analysis for \CoP{} that reveal the intrinsic nature of the quantum fluctuations in the spin excitation spectra of the long-range magnetically ordered phases. In particular, we observe significant weight transferred from the 1-magnon excitation to the 2-magnon continuum, and strong magnon decay when they satisfy the kinematic condition. The broadening of the spin wave is strong especially at zero field where the ground state is a long-range ordered ``Y'' phase, where only a continuum of excitation is seen. This observation challenges the conventional practice in the field that often associate such continuum as signature of the QSL -- as we have shown, spin excitation continuum is also fully compatible with a long-range magnetically ordered state with strong quantum fluctuations.

\bibliographystyle{naturemag}
\bibliography{refs}

\begin{thebibliography}{10}
\expandafter\ifx\csname url\endcsname\relax
  \def\url#1{\texttt{#1}}\fi
\expandafter\ifx\csname urlprefix\endcsname\relax\def\urlprefix{URL }\fi
\providecommand{\bibinfo}[2]{#2}
\providecommand{\eprint}[2][]{\url{#2}}

\bibitem{AndersonPW1973}
\bibinfo{author}{Anderson, P.~W.}
\newblock \bibinfo{title}{Resonating valence bonds: {{A}} new kind of
  insulator?}
\newblock \emph{\bibinfo{journal}{Mater. Res. Bull.}}
  \textbf{\bibinfo{volume}{8}}, \bibinfo{pages}{153--160}
  (\bibinfo{year}{1973}).
\newblock
  \urlprefix\url{https://www.sciencedirect.com/science/article/pii/0025540873901670}.

\bibitem{KitaevAY2003}
\bibinfo{author}{Kitaev, A.~{\relax Yu}.}
\newblock \bibinfo{title}{Fault-tolerant quantum computation by anyons}.
\newblock \emph{\bibinfo{journal}{Ann. Phys.}} \textbf{\bibinfo{volume}{303}},
  \bibinfo{pages}{2--30} (\bibinfo{year}{2003}).
\newblock
  \urlprefix\url{https://www.sciencedirect.com/science/article/pii/S0003491602000180}.

\bibitem{KitaevA2006}
\bibinfo{author}{Kitaev, A.}
\newblock \bibinfo{title}{Anyons in an exactly solved model and beyond}.
\newblock \emph{\bibinfo{journal}{Ann. Phys.}} \textbf{\bibinfo{volume}{321}},
  \bibinfo{pages}{2--111} (\bibinfo{year}{2006}).
\newblock
  \urlprefix\url{https://www.sciencedirect.com/science/article/pii/S0003491605002381}.

\bibitem{AndersonPW1987}
\bibinfo{author}{Anderson, P.~W.}
\newblock \bibinfo{title}{The {{Resonating Valence Bond State}} in
  {{La}}{$_{2}$}{{CuO}}{$_4$} and {{Superconductivity}}}.
\newblock \emph{\bibinfo{journal}{Science}} \textbf{\bibinfo{volume}{235}},
  \bibinfo{pages}{1196--1198} (\bibinfo{year}{1987}).
\newblock
  \urlprefix\url{https://www.science.org/doi/10.1126/science.235.4793.1196}.

\bibitem{LeePA2006_RMP}
\bibinfo{author}{Lee, P.~A.}, \bibinfo{author}{Nagaosa, N.} \&
  \bibinfo{author}{Wen, X.-G.}
\newblock \bibinfo{title}{Doping a {{Mott}} insulator: {{Physics}} of
  high-temperature superconductivity}.
\newblock \emph{\bibinfo{journal}{Rev. Mod. Phys.}}
  \textbf{\bibinfo{volume}{78}}, \bibinfo{pages}{17--85}
  (\bibinfo{year}{2006}).
\newblock \urlprefix\url{https://link.aps.org/doi/10.1103/RevModPhys.78.17}.

\bibitem{BalentsL2010_review}
\bibinfo{author}{Balents, L.}
\newblock \bibinfo{title}{Spin liquids in frustrated magnets}.
\newblock \emph{\bibinfo{journal}{Nature}} \textbf{\bibinfo{volume}{464}},
  \bibinfo{pages}{199--208} (\bibinfo{year}{2010}).
\newblock \urlprefix\url{https://www.nature.com/articles/nature08917}.

\bibitem{ZhouY2017_RMP}
\bibinfo{author}{Zhou, Y.}, \bibinfo{author}{Kanoda, K.} \&
  \bibinfo{author}{Ng, T.-K.}
\newblock \bibinfo{title}{Quantum spin liquid states}.
\newblock \emph{\bibinfo{journal}{Rev. Mod. Phys.}}
  \textbf{\bibinfo{volume}{89}}, \bibinfo{pages}{025003}
  (\bibinfo{year}{2017}).
\newblock
  \urlprefix\url{https://link.aps.org/doi/10.1103/RevModPhys.89.025003}.

\bibitem{WenJ2019_review}
\bibinfo{author}{Wen, J.}, \bibinfo{author}{Yu, S.-L.}, \bibinfo{author}{Li,
  S.}, \bibinfo{author}{Yu, W.} \& \bibinfo{author}{Li, J.-X.}
\newblock \bibinfo{title}{Experimental identification of quantum spin liquids}.
\newblock \emph{\bibinfo{journal}{npj Quantum Mater.}}
  \textbf{\bibinfo{volume}{4}}, \bibinfo{pages}{1--9} (\bibinfo{year}{2019}).
\newblock \urlprefix\url{https://www.nature.com/articles/s41535-019-0151-6}.

\bibitem{BroholmC2020_review}
\bibinfo{author}{Broholm, C.} \emph{et~al.}
\newblock \bibinfo{title}{Quantum spin liquids}.
\newblock \emph{\bibinfo{journal}{Science}} \textbf{\bibinfo{volume}{367}},
  \bibinfo{pages}{eaay0668} (\bibinfo{year}{2020}).
\newblock \urlprefix\url{https://www.science.org/doi/10.1126/science.aay0668}.

\bibitem{ShimizuY2003}
\bibinfo{author}{Shimizu, Y.}, \bibinfo{author}{Miyagawa, K.},
  \bibinfo{author}{Kanoda, K.}, \bibinfo{author}{Maesato, M.} \&
  \bibinfo{author}{Saito, G.}
\newblock \bibinfo{title}{Spin {{Liquid State}} in an {{Organic Mott
  Insulator}} with a {{Triangular Lattice}}}.
\newblock \emph{\bibinfo{journal}{Phys. Rev. Lett.}}
  \textbf{\bibinfo{volume}{91}}, \bibinfo{pages}{107001}
  (\bibinfo{year}{2003}).
\newblock
  \urlprefix\url{https://link.aps.org/doi/10.1103/PhysRevLett.91.107001}.

\bibitem{ItouT2008}
\bibinfo{author}{Itou, T.}, \bibinfo{author}{Oyamada, A.},
  \bibinfo{author}{Maegawa, S.}, \bibinfo{author}{Tamura, M.} \&
  \bibinfo{author}{Kato, R.}
\newblock \bibinfo{title}{Quantum spin liquid in the spin-1/2 triangular
  antiferromagnet {{EtMe}}{$_{3}$}{{Sb}}[{{Pd}}(dmit){$_2$}]{$_{2}$}}.
\newblock \emph{\bibinfo{journal}{Phys. Rev. B}} \textbf{\bibinfo{volume}{77}},
  \bibinfo{pages}{104413} (\bibinfo{year}{2008}).
\newblock \urlprefix\url{https://link.aps.org/doi/10.1103/PhysRevB.77.104413}.

\bibitem{IsonoT2014}
\bibinfo{author}{Isono, T.} \emph{et~al.}
\newblock \bibinfo{title}{Gapless {{Quantum Spin Liquid}} in an {{Organic
  Spin-1}}/2 {{Triangular-Lattice}}
  {$\kappa$}-{{H}}{$_3$}({{Cat-EDT-TTF}}){$_{2}$}}.
\newblock \emph{\bibinfo{journal}{Phys. Rev. Lett.}}
  \textbf{\bibinfo{volume}{112}}, \bibinfo{pages}{177201}
  (\bibinfo{year}{2014}).
\newblock
  \urlprefix\url{https://link.aps.org/doi/10.1103/PhysRevLett.112.177201}.

\bibitem{HanTH2012}
\bibinfo{author}{Han, T.-H.} \emph{et~al.}
\newblock \bibinfo{title}{Fractionalized excitations in the spin-liquid state
  of a kagome-lattice antiferromagnet}.
\newblock \emph{\bibinfo{journal}{Nature}} \textbf{\bibinfo{volume}{492}},
  \bibinfo{pages}{406--410} (\bibinfo{year}{2012}).
\newblock \urlprefix\url{https://www.nature.com/articles/nature11659}.

\bibitem{XuS2023}
\bibinfo{author}{Xu, S.} \emph{et~al.}
\newblock \bibinfo{title}{Realization of {{U}}(1) {{Dirac Quantum Spin Liquid}}
  in {{YbZn}}{$_{2}$}{{GaO}}{$_{5}$}} (\bibinfo{year}{2023}).
\newblock \urlprefix\url{http://arxiv.org/abs/2305.20040}.

\bibitem{LiuW2018}
\bibinfo{author}{Liu, W.} \emph{et~al.}
\newblock \bibinfo{title}{Rare-{{Earth Chalcogenides}}: {{A Large Family}} of
  {{Triangular Lattice Spin Liquid Candidates}}}.
\newblock \emph{\bibinfo{journal}{Chinese Phys. Lett.}}
  \textbf{\bibinfo{volume}{35}}, \bibinfo{pages}{117501}
  (\bibinfo{year}{2018}).
\newblock \urlprefix\url{https://doi.org/10.1088/0256-307x/35/11/117501}.

\bibitem{BaenitzM2018}
\bibinfo{author}{Baenitz, M.} \emph{et~al.}
\newblock \bibinfo{title}{{{NaYbS}}{$_2$}: {{A}} planar spin-1/2
  triangular-lattice magnet and putative spin liquid}.
\newblock \emph{\bibinfo{journal}{Phys. Rev. B}} \textbf{\bibinfo{volume}{98}},
  \bibinfo{pages}{220409} (\bibinfo{year}{2018}).
\newblock \urlprefix\url{https://link.aps.org/doi/10.1103/PhysRevB.98.220409}.

\bibitem{BordelonMM2019}
\bibinfo{author}{Bordelon, M.~M.} \emph{et~al.}
\newblock \bibinfo{title}{Field-tunable quantum disordered ground state in the
  triangular-lattice antiferromagnet {{NaYbO}}{$_{2}$}}.
\newblock \emph{\bibinfo{journal}{Nat. Phys.}} \textbf{\bibinfo{volume}{15}},
  \bibinfo{pages}{1058--1064} (\bibinfo{year}{2019}).
\newblock \urlprefix\url{https://www.nature.com/articles/s41567-019-0594-5}.

\bibitem{DaiPL2021}
\bibinfo{author}{Dai, P.-L.} \emph{et~al.}
\newblock \bibinfo{title}{Spinon {{Fermi Surface Spin Liquid}} in a
  {{Triangular Lattice Antiferromagnet NaYbSe}}{$_{2}$}}.
\newblock \emph{\bibinfo{journal}{Phys. Rev. X}} \textbf{\bibinfo{volume}{11}},
  \bibinfo{pages}{021044} (\bibinfo{year}{2021}).
\newblock \urlprefix\url{https://link.aps.org/doi/10.1103/PhysRevX.11.021044}.

\bibitem{ScheieAO2024}
\bibinfo{author}{Scheie, A.~O.} \emph{et~al.}
\newblock \bibinfo{title}{Nonlinear magnons and exchange {{Hamiltonians}} of
  the delafossite proximate quantum spin liquid candidates {{KYbSe}}{$_2$} and
  {{NaYbSe}}{$_{2}$}}.
\newblock \emph{\bibinfo{journal}{Phys. Rev. B}}
  \textbf{\bibinfo{volume}{109}}, \bibinfo{pages}{014425}
  (\bibinfo{year}{2024}).
\newblock \urlprefix\url{https://link.aps.org/doi/10.1103/PhysRevB.109.014425}.

\bibitem{LawKT2017}
\bibinfo{author}{Law, K.~T.} \& \bibinfo{author}{Lee, P.~A.}
\newblock \bibinfo{title}{{{1T-TaS}}{$_2$} as a quantum spin liquid}.
\newblock \emph{\bibinfo{journal}{Proc. Natl. Acad. Sci. U.S.A.}}
  \textbf{\bibinfo{volume}{114}}, \bibinfo{pages}{6996--7000}
  (\bibinfo{year}{2017}).
\newblock
  \urlprefix\url{https://www.pnas.org/doi/full/10.1073/pnas.1706769114}.

\bibitem{WangZ2017_RuCl3}
\bibinfo{author}{Wang, Z.} \emph{et~al.}
\newblock \bibinfo{title}{Magnetic {{Excitations}} and {{Continuum}} of a
  {{Possibly Field-Induced Quantum Spin Liquid}} in
  {$\alpha$}-{{RuCl}}{$_{3}$}}.
\newblock \emph{\bibinfo{journal}{Phys. Rev. Lett.}}
  \textbf{\bibinfo{volume}{119}}, \bibinfo{pages}{227202}
  (\bibinfo{year}{2017}).
\newblock
  \urlprefix\url{https://link.aps.org/doi/10.1103/PhysRevLett.119.227202}.

\bibitem{ZhengJ2017}
\bibinfo{author}{Zheng, J.} \emph{et~al.}
\newblock \bibinfo{title}{Gapless {{Spin Excitations}} in the {{Field-Induced
  Quantum Spin Liquid Phase}} of {$\alpha$}-{{RuCl}}{$_{3}$}}.
\newblock \emph{\bibinfo{journal}{Phys. Rev. Lett.}}
  \textbf{\bibinfo{volume}{119}}, \bibinfo{pages}{227208}
  (\bibinfo{year}{2017}).
\newblock
  \urlprefix\url{https://link.aps.org/doi/10.1103/PhysRevLett.119.227208}.

\bibitem{HalloranT2023}
\bibinfo{author}{Halloran, T.} \emph{et~al.}
\newblock \bibinfo{title}{Geometrical frustration versus {{Kitaev}}
  interactions in {{BaCo}}{$_2$}({{AsO}}{$_4$}){$_{2}$}}.
\newblock \emph{\bibinfo{journal}{Proc. Natl. Acad. Sci. U.S.A.}}
  \textbf{\bibinfo{volume}{120}}, \bibinfo{pages}{e2215509119}
  (\bibinfo{year}{2023}).
\newblock \urlprefix\url{https://www.pnas.org/doi/10.1073/pnas.2215509119}.

\bibitem{BalzC2016}
\bibinfo{author}{Balz, C.} \emph{et~al.}
\newblock \bibinfo{title}{Physical realization of a quantum spin liquid based
  on a complex frustration mechanism}.
\newblock \emph{\bibinfo{journal}{Nat. Phys.}} \textbf{\bibinfo{volume}{12}},
  \bibinfo{pages}{942--949} (\bibinfo{year}{2016}).
\newblock \urlprefix\url{https://www.nature.com/articles/nphys3826}.

\bibitem{OkamotoY2007}
\bibinfo{author}{Okamoto, Y.}, \bibinfo{author}{Nohara, M.},
  \bibinfo{author}{{Aruga-Katori}, H.} \& \bibinfo{author}{Takagi, H.}
\newblock \bibinfo{title}{Spin-{{Liquid State}} in the {{{\emph{S}}}}=1/2
  {{Hyperkagome Antiferromagnet Na}}{$_{4}$}{{Ir}}{$_{3}$}{{O}}{$_{8}$}}.
\newblock \emph{\bibinfo{journal}{Phys. Rev. Lett.}}
  \textbf{\bibinfo{volume}{99}}, \bibinfo{pages}{137207}
  (\bibinfo{year}{2007}).
\newblock
  \urlprefix\url{https://link.aps.org/doi/10.1103/PhysRevLett.99.137207}.

\bibitem{ZivkovicI2021}
\bibinfo{author}{{\v Z}ivkovi{\'c}, I.} \emph{et~al.}
\newblock \bibinfo{title}{Magnetic {{Field Induced Quantum Spin Liquid}} in the
  {{Two Coupled Trillium Lattices}} of
  {{K}}{$_{2}$}{{Ni}}{$_2$}({{SO}}{$_4$}){$_{3}$}}.
\newblock \emph{\bibinfo{journal}{Phys. Rev. Lett.}}
  \textbf{\bibinfo{volume}{127}}, \bibinfo{pages}{157204}
  (\bibinfo{year}{2021}).
\newblock
  \urlprefix\url{https://link.aps.org/doi/10.1103/PhysRevLett.127.157204}.

\bibitem{ZhuW2019}
\bibinfo{author}{Zhu, W.}, \bibinfo{author}{Gong, S.-s.} \&
  \bibinfo{author}{Sheng, D.~N.}
\newblock \bibinfo{title}{Identifying spinon excitations from dynamic structure
  factor of spin-1/2 {{Heisenberg}} antiferromagnet on the {{Kagome}} lattice}.
\newblock \emph{\bibinfo{journal}{Proc. Natl. Acad. Sci. U.S.A.}}
  \textbf{\bibinfo{volume}{116}}, \bibinfo{pages}{5437--5441}
  (\bibinfo{year}{2019}).
\newblock \urlprefix\url{https://www.pnas.org/doi/10.1073/pnas.1807840116}.

\bibitem{ZhongR2019}
\bibinfo{author}{Zhong, R.}, \bibinfo{author}{Guo, S.}, \bibinfo{author}{Xu,
  G.}, \bibinfo{author}{Xu, Z.} \& \bibinfo{author}{Cava, R.~J.}
\newblock \bibinfo{title}{Strong quantum fluctuations in a quantum spin liquid
  candidate with a {{Co-based}} triangular lattice}.
\newblock \emph{\bibinfo{journal}{Proc. Natl. Acad. Sci. U.S.A.}}
  \textbf{\bibinfo{volume}{116}}, \bibinfo{pages}{14505--14510}
  (\bibinfo{year}{2019}).
\newblock \urlprefix\url{https://www.pnas.org/doi/10.1073/pnas.1906483116}.

\bibitem{LeeS2021}
\bibinfo{author}{Lee, S.} \emph{et~al.}
\newblock \bibinfo{title}{Temporal and field evolution of spin excitations in
  the disorder-free triangular antiferromagnet
  {{Na}}{\textsubscript{2}}{{BaCo}}({{PO}}{\textsubscript{4}}){\textsubscript{2}}}.
\newblock \emph{\bibinfo{journal}{Phys. Rev. B}}
  \textbf{\bibinfo{volume}{103}}, \bibinfo{pages}{024413}
  (\bibinfo{year}{2021}).
\newblock \urlprefix\url{https://link.aps.org/doi/10.1103/PhysRevB.103.024413}.

\bibitem{LiN2020}
\bibinfo{author}{Li, N.} \emph{et~al.}
\newblock \bibinfo{title}{Possible itinerant excitations and quantum spin state
  transitions in the effective spin-1/2 triangular-lattice antiferromagnet
  {{Na}}{$_{2}$}{{BaCo}}({{PO}}{$_4$}){$_{2}$}}.
\newblock \emph{\bibinfo{journal}{Nat. Commun.}} \textbf{\bibinfo{volume}{11}},
  \bibinfo{pages}{4216} (\bibinfo{year}{2020}).
\newblock \urlprefix\url{https://www.nature.com/articles/s41467-020-18041-3}.

\bibitem{ShengJ2022_NaBaCoPO}
\bibinfo{author}{Sheng, J.} \emph{et~al.}
\newblock \bibinfo{title}{Two-dimensional quantum universality in the spin-1/2
  triangular-lattice quantum antiferromagnet
  {{Na}}{$_{2}$}{{BaCo}}({{PO}}{$_4$}){$_{2}$}}.
\newblock \emph{\bibinfo{journal}{Proc. Natl. Acad. Sci. U.S.A.}}
  \textbf{\bibinfo{volume}{119}}, \bibinfo{pages}{e2211193119}
  (\bibinfo{year}{2022}).
\newblock \urlprefix\url{https://www.pnas.org/doi/abs/10.1073/pnas.2211193119}.

\bibitem{GaoY2022_NaBaCoPO}
\bibinfo{author}{Gao, Y.} \emph{et~al.}
\newblock \bibinfo{title}{Spin supersolidity in nearly ideal easy-axis
  triangular quantum antiferromagnet
  {{Na}}{$_{2}$}{{BaCo}}({{PO}}{$_4$}){$_{2}$}}.
\newblock \emph{\bibinfo{journal}{npj Quantum Mater.}}
  \textbf{\bibinfo{volume}{7}}, \bibinfo{pages}{1--8} (\bibinfo{year}{2022}).
\newblock \urlprefix\url{https://www.nature.com/articles/s41535-022-00500-3}.

\bibitem{ZapfV2014_RMP}
\bibinfo{author}{Zapf, V.}, \bibinfo{author}{Jaime, M.} \&
  \bibinfo{author}{Batista, C.~D.}
\newblock \bibinfo{title}{Bose-{{Einstein}} condensation in quantum magnets}.
\newblock \emph{\bibinfo{journal}{Rev. Mod. Phys.}}
  \textbf{\bibinfo{volume}{86}}, \bibinfo{pages}{563--614}
  (\bibinfo{year}{2014}).
\newblock \urlprefix\url{https://link.aps.org/doi/10.1103/RevModPhys.86.563}.

\bibitem{AndreevAF1969}
\bibinfo{author}{Andreev, A.~F.} \& \bibinfo{author}{Lifshitz, I.~M.}
\newblock \bibinfo{title}{Quantum {{Theory}} of {{Defects}} in {{Crystals}}}.
\newblock \emph{\bibinfo{journal}{Zh. Eksp. Teor. Fiz.}}
  \textbf{\bibinfo{volume}{56}}, \bibinfo{pages}{2057--2068}
  (\bibinfo{year}{1969}).
\newblock \bibinfo{note}{[Sov. Phys. JETP 29, 1107 (1969)]}.

\bibitem{ChesterGV1970}
\bibinfo{author}{Chester, G.~V.}
\newblock \bibinfo{title}{Speculations on {{Bose-Einstein Condensation}} and
  {{Quantum Crystals}}}.
\newblock \emph{\bibinfo{journal}{Phys. Rev. A}} \textbf{\bibinfo{volume}{2}},
  \bibinfo{pages}{256--258} (\bibinfo{year}{1970}).
\newblock \urlprefix\url{https://link.aps.org/doi/10.1103/PhysRevA.2.256}.

\bibitem{LeggettAJ1970}
\bibinfo{author}{Leggett, A.~J.}
\newblock \bibinfo{title}{Can a {{Solid Be}} "{{Superfluid}}"?}
\newblock \emph{\bibinfo{journal}{Phys. Rev. Lett.}}
  \textbf{\bibinfo{volume}{25}}, \bibinfo{pages}{1543--1546}
  (\bibinfo{year}{1970}).
\newblock \urlprefix\url{https://link.aps.org/doi/10.1103/PhysRevLett.25.1543}.

\bibitem{YamamotoD2014}
\bibinfo{author}{Yamamoto, D.}, \bibinfo{author}{Marmorini, G.} \&
  \bibinfo{author}{Danshita, I.}
\newblock \bibinfo{title}{Quantum {{Phase Diagram}} of the
  {{Triangular-Lattice}} {{{\emph{XXZ}}}} {{Model}} in a {{Magnetic Field}}}.
\newblock \emph{\bibinfo{journal}{Phys. Rev. Lett.}}
  \textbf{\bibinfo{volume}{112}}, \bibinfo{pages}{127203}
  (\bibinfo{year}{2014}).
\newblock
  \urlprefix\url{https://link.aps.org/doi/10.1103/PhysRevLett.112.127203}.

\bibitem{SellmannD2015}
\bibinfo{author}{Sellmann, D.}, \bibinfo{author}{Zhang, X.-F.} \&
  \bibinfo{author}{Eggert, S.}
\newblock \bibinfo{title}{Phase diagram of the antiferromagnetic {{XXZ}} model
  on the triangular lattice}.
\newblock \emph{\bibinfo{journal}{Phys. Rev. B}} \textbf{\bibinfo{volume}{91}},
  \bibinfo{pages}{081104} (\bibinfo{year}{2015}).
\newblock \urlprefix\url{https://link.aps.org/doi/10.1103/PhysRevB.91.081104}.

\bibitem{KuhnerTD1999}
\bibinfo{author}{K{\"u}hner, T.~D.} \& \bibinfo{author}{White, S.~R.}
\newblock \bibinfo{title}{Dynamical correlation functions using the density
  matrix renormalization group}.
\newblock \emph{\bibinfo{journal}{Phys. Rev. B}} \textbf{\bibinfo{volume}{60}},
  \bibinfo{pages}{335--343} (\bibinfo{year}{1999}).
\newblock \urlprefix\url{https://link.aps.org/doi/10.1103/PhysRevB.60.335}.

\bibitem{JeckelmannE2002}
\bibinfo{author}{Jeckelmann, E.}
\newblock \bibinfo{title}{Dynamical density-matrix renormalization-group
  method}.
\newblock \emph{\bibinfo{journal}{Phys. Rev. B}} \textbf{\bibinfo{volume}{66}},
  \bibinfo{pages}{045114} (\bibinfo{year}{2002}).
\newblock \urlprefix\url{https://link.aps.org/doi/10.1103/PhysRevB.66.045114}.

\bibitem{ZhitomirskyME2013_RMP}
\bibinfo{author}{Zhitomirsky, M.~E.} \& \bibinfo{author}{Chernyshev, A.~L.}
\newblock \bibinfo{title}{Colloquium: {{Spontaneous}} magnon decays}.
\newblock \emph{\bibinfo{journal}{Rev. Mod. Phys.}}
  \textbf{\bibinfo{volume}{85}}, \bibinfo{pages}{219--242}
  (\bibinfo{year}{2013}).
\newblock \urlprefix\url{https://link.aps.org/doi/10.1103/RevModPhys.85.219}.

\bibitem{ZhuZ2015}
\bibinfo{author}{Zhu, Z.} \& \bibinfo{author}{White, S.~R.}
\newblock \bibinfo{title}{Spin liquid phase of the {{{\emph{S}}}}=1/2
  {{{\emph{J}}}}{\emph{{$_{1}$}}}-{{{\emph{J}}}}{\emph{{$_{2}$}}}
  {{Heisenberg}} model on the triangular lattice}.
\newblock \emph{\bibinfo{journal}{Phys. Rev. B}} \textbf{\bibinfo{volume}{92}},
  \bibinfo{pages}{041105(R)} (\bibinfo{year}{2015}).
\newblock \urlprefix\url{https://link.aps.org/doi/10.1103/PhysRevB.92.041105}.

\bibitem{HuWJ2015}
\bibinfo{author}{Hu, W.-J.}, \bibinfo{author}{Gong, S.-S.},
  \bibinfo{author}{Zhu, W.} \& \bibinfo{author}{Sheng, D.~N.}
\newblock \bibinfo{title}{Competing spin-liquid states in the spin-1/2
  {{Heisenberg}} model on the triangular lattice}.
\newblock \emph{\bibinfo{journal}{Phys. Rev. B}} \textbf{\bibinfo{volume}{92}},
  \bibinfo{pages}{140403(R)} (\bibinfo{year}{2015}).
\newblock \urlprefix\url{https://link.aps.org/doi/10.1103/PhysRevB.92.140403}.

\bibitem{GhioldiEA2018}
\bibinfo{author}{Ghioldi, E.~A.} \emph{et~al.}
\newblock \bibinfo{title}{Dynamical structure factor of the triangular
  antiferromagnet: {{Schwinger}} boson theory beyond mean field}.
\newblock \emph{\bibinfo{journal}{Phys. Rev. B}} \textbf{\bibinfo{volume}{98}},
  \bibinfo{pages}{184403} (\bibinfo{year}{2018}).
\newblock \urlprefix\url{https://link.aps.org/doi/10.1103/PhysRevB.98.184403}.

\bibitem{JiaH2023}
\bibinfo{author}{Jia, H.}, \bibinfo{author}{Ma, B.}, \bibinfo{author}{Wang, Z.}
  \& \bibinfo{author}{Chen, G.}
\newblock \bibinfo{title}{Quantum {{Spin Supersolid}} as a precursory {{Dirac
  Spin Liquid}} in a {{Triangular Lattice Antiferromagnet}}}
  (\bibinfo{year}{2023}).
\newblock \urlprefix\url{http://arxiv.org/abs/2304.11716}.

\bibitem{XieT2023}
\bibinfo{author}{Xie, T.} \emph{et~al.}
\newblock \bibinfo{title}{Complete field-induced spectral response of the
  spin-1/2 triangular-lattice antiferromagnet {{CsYbSe}}{$_{2}$}}.
\newblock \emph{\bibinfo{journal}{npj Quantum Mater.}}
  \textbf{\bibinfo{volume}{8}}, \bibinfo{pages}{1--9} (\bibinfo{year}{2023}).
\newblock \urlprefix\url{https://www.nature.com/articles/s41535-023-00580-9}.

\bibitem{ScheieAO2023}
\bibinfo{author}{Scheie, A.~O.} \emph{et~al.}
\newblock \bibinfo{title}{Proximate spin liquid and fractionalization in the
  triangular antiferromagnet {{KYbSe}}{$_{2}$}}.
\newblock \emph{\bibinfo{journal}{Nat. Phys.}} \textbf{\bibinfo{volume}{20}},
  \bibinfo{pages}{74--81} (\bibinfo{year}{2023}).
\newblock \urlprefix\url{https://www.nature.com/articles/s41567-023-02259-1}.

\bibitem{NakajimaK2011}
\bibinfo{author}{Nakajima, K.} \emph{et~al.}
\newblock \bibinfo{title}{{{AMATERAS}}: {{A Cold-Neutron Disk Chopper
  Spectrometer}}}.
\newblock \emph{\bibinfo{journal}{J. Phys. Soc. Jpn.}}
  \textbf{\bibinfo{volume}{80}}, \bibinfo{pages}{SB028} (\bibinfo{year}{2011}).
\newblock \urlprefix\url{https://journals.jps.jp/doi/10.1143/JPSJS.80SB.SB028}.

\bibitem{InamuraY2013}
\bibinfo{author}{Inamura, Y.}, \bibinfo{author}{Nakatani, T.},
  \bibinfo{author}{Suzuki, J.} \& \bibinfo{author}{Otomo, T.}
\newblock \bibinfo{title}{Development {{Status}} of {{Software}}
  ``{{Utsusemi}}'' for {{Chopper Spectrometers}} at {{MLF}}, {{J-PARC}}}.
\newblock \emph{\bibinfo{journal}{J. Phys. Soc. Jpn.}}
  \textbf{\bibinfo{volume}{82}}, \bibinfo{pages}{SA031} (\bibinfo{year}{2013}).
\newblock \urlprefix\url{https://journals.jps.jp/doi/10.7566/JPSJS.82SA.SA031}.

\bibitem{AzuahRT2009}
\bibinfo{author}{Azuah, R.~T.} \emph{et~al.}
\newblock \bibinfo{title}{Dave: {{A}} compressive software suite for the
  reduction, visualization, and analysis of low energy neutron spectroscopic
  data}.
\newblock \emph{\bibinfo{journal}{J. Res. Natl. Inst. Stan. Technol.}}
  \textbf{\bibinfo{volume}{114}}, \bibinfo{pages}{341--358}
  (\bibinfo{year}{2009}).
\newblock \urlprefix\url{http://dx.doi.org/10.6028/jres.114.025}.

\end{thebibliography}

\newpage
\textbf{Methods}

{\bf Inelastic neutron scattering.}
Neutron scattering experiments were conducted at the cold-neutron disk chopper spectrometer, AMATERAS (BL14 beamline), with fixed incident energy $E_i=\qty{2.63}{meV}$ (energy resolution is about $\qty{0.047}{meV}$) at the Materials and Life Science Experimental Facility (MLF), J-PARC~\cite{NakajimaK2011}, and the time-of-flight cold neutron spectrometer, PELICAN, at the OPAL reactor, ANSTO, with a fixed incident energy $E_i=\qty{3.71}{meV}$ (energy resolution is about $\qty{0.13}{meV}$). 
The INS results presented in the main text are from AMATERAS, and the ones from PELICAN are shown in the Supplementary Materials for comparison.
Hundreds of single crystal \CoP{} were co-aligned on the oxygen-free copper sheets for the INS experiments, with totaling approximately 3 grams in mass. The samples were cooled using a dilution refrigerator insert in a \qty{7}{T} magnet on both spectrometers, with the magnetic field applied along the $c$-axis. The INS data were collected at temperatures of $\qty{60}{mK}$, $\qty{450}{mK}$ with different magnetic fields and processed using the freely available Utsusemi~\cite{InamuraY2013} and Dave software tool~\cite{AzuahRT2009}. 

{\bf Linear spin wave.}
The semiclassical ground states of the TL XXZ model \eqref{eq:xxz} contain 3 sublattices in the magnetic unit cell for $B<B_\text{s}$, which we denote as S\{$\vec{s}_1$, $\vec{s}_2$, $\vec{s}_3$\} where $|\vec{s}_i|=1$ and $S\equiv 1/2$. Such $\sqrt{3}\times \sqrt{3}$ magnetic structures form a superlattice with real-space basis $\bm{A}_1 = 2 \bm{a}_1 + \bm{a}_2$ and $\bm{A}_2 = \bm{a}_1 + 2 \bm{a}_2$, where \{$\bm{a}_1$, $\bm{a}_2$\} is the basis of the original lattice. 

Minimization of the classical energy at $T=0$ with respect to \{$\vec{s}_1$, $\vec{s}_2$, $\vec{s}_3$\} indeed reveals 3 phases below saturation. At intermediate field, we have $-\vec{s}_1=\vec{s}_2=\vec{s}_3=(0,0,1)$, namely the UUD phase. 

At low field, the optimal classical spin configuration is the ``Y'' phase:
\begin{equation}
\vec{s}_1=(0,0,-1), \quad \vec{s}_2 = (\sin \theta, 0, \cos \theta),\quad \vec{s}_3 = (-\sin \theta, 0, \cos \theta),
\end{equation}
up to a global $U(1)$ rotation around the $z$-axis. The optimal value is 
\begin{equation}\cos \theta = \frac{\Delta + g_c \mu_B B/(3JS)}{\Delta + 1}.
\end{equation}

The high-field ``V'' phase is:
\begin{equation}
\vec{s}_1=(-\sin \theta_1, 0, \cos \theta_1),\quad \vec{s}_2 = \vec{s}_3 = (\sin \theta_2, 0, \cos \theta_2),
\end{equation}
up to a global $U(1)$ rotation around the $z$-axis. The optimal values of \{$\theta_1$ $\theta_2$\} are obtained by numerically minimizing the classical energy in this paper.

To perform the LSW, we first rotate to a local frame where the spins point to the $\hat{z}\equiv (0,0,1)$ direction:
\begin{equation}
\vec{s}_i = R_i \hat{z},
\end{equation}
where $R_i$ defines the SO(3) rotation matrix on each site. Accordingly, the spin operators transform as
\begin{equation}
\bm{S}_i = R_i \tilde{\bm{S}}_i .
\end{equation}

By representing the spin operators $\tilde{\bm{S}}_i$ with Holstein-Primakoff bosons:
\begin{subequations}
\begin{align}
\tilde{S}_i^+ &= \sqrt{2S - b_i^\dagger b_i} b_i,\\
\tilde{S}_i^- &= b_i^\dagger \sqrt{2S - b_i^\dagger b_i},\\
\tilde{S}_i^z &= S- b_i^\dagger b_i,
\end{align}
\end{subequations}
and keep up to quadratic order in the bosonic operators, we obtain the following Hamiltonian in Fouier space:
\begin{equation}
\mathcal{H} \approx \sum_{\tilde{\bm{k}}}^\prime \Psi_{\tilde{\bm{k}}}^\dagger H_\text{LSW}(\tilde{\bm{k}}) \Psi_{\tilde{\bm{k}}} + C,
\end{equation}
where $C$ is a constant, $\Psi_{\tilde{\bm{k}}}\equiv \left( b_{\tilde{\bm{k}},\bm{d}_1},b_{\tilde{\bm{k}},\bm{d}_2},b_{\tilde{\bm{k}},\bm{d}_3},b_{-\tilde{\bm{k}},\bm{d}_1}^\dagger, b_{-\tilde{\bm{k}},\bm{d}_2}^\dagger,b_{-\tilde{\bm{k}},\bm{d}_3}^\dagger \right)^T$, $H_\text{LSW}(\tilde{\bm{k}})$ is a $6\times 6$ Hermitian matrix, and the prime on the summation denotes that we are only summing over half of the folded Brillouin zone. Note that the presence of a superlattice requires a compatible Fourier transformation:
\begin{equation}
b_i \equiv b_{\tilde{\bm{r}}+\bm{d}} = \sqrt{\frac{3}{N}} \sum_{\tilde{\bm{k}}} e^{\iu \tilde{\bm{k}} \cdot \tilde{\bm{r}}} b_{\tilde{\bm{k}},\bm{d}},
\end{equation}
where $\tilde{\bm{r}} = m\bm{A}_1+n\bm{A}_2$ are positions of the superlattice, $\bm{d}=\{\bm{d}_1,\bm{d}_2,\bm{d}_3\}$ are positions of the 3 sublattices, and $N\rightarrow \infty$ is the total number of lattice sites. Note that $\tilde{\bm{k}}$ is related to the original momentum $\bm{k}$ by $\bm{k}=\tilde{\bm{k}} + \bm{K}$, where $\bm{K}$ denotes lattice points of the reciprocal superlattice.


The LSW Hamiltonian can be diagonalized by the Bogoliubov transformation:
\begin{align}
\Psi_{\tilde{\bm{k}}} &= V_{\tilde{\bm{k}}} \tilde{\Psi}_{\tilde{\bm{k}}},\\
V_{\tilde{\bm{k}}}^\dagger H_\text{LSW}(\tilde{\bm{k}}) V_{\tilde{\bm{k}}} &= \text{diag} \{ \omega_{\tilde{\bm{k}},3}, \omega_{\tilde{\bm{k}},2}, \omega_{\tilde{\bm{k}},1}, \omega_{-\tilde{\bm{k}},1}, \omega_{-\tilde{\bm{k}},2}, \omega_{-\tilde{\bm{k}},3} \}.
\end{align}

The dynamic spin structure factor including both the transverse (1-magnon) and longitudinal (2-magnon continuum) modes:
\begin{align}
\mathcal{S}^{ab}(\bm{k},\omega) &= \pi S \sum_{j=1}^3 A_j^a(\bm{k}) \left[ A_j^b (\bm{k})\right]^* \delta \left( \omega - \omega_{\tilde{\bm{k}},j} \right) \nonumber \\
& \quad + 2\pi \int \frac{\mathrm{d}\tilde{\bm{q}}}{\mathcal{A}_\text{BZ}} \sum_{j_1, j_2=1}^3 A_{j_1j_2}^a (\bm{k},\tilde{\bm{q}}) \left[ A_{j_1j_2}^b (\bm{k},\tilde{\bm{q}}) \right]^* \delta (\omega - \omega_{-\tilde{\bm{q}},j_1} - \omega_{\widetilde{\bm{q}+\bm{k}},j_2}),
\end{align}
where $\mathcal{A}_\text{BZ}$ is the area of the folded Brillouin zone, and
\begin{subequations}
\begin{align}
\vec{A}_j (\bm{k}) &\equiv \frac{1}{\sqrt{3}} \sum_{l=1}^3 e^{-\iu \bm{k} \cdot \bm{d}_l} R_l 
\begin{pmatrix}\left(V_{\tilde{\bm{k}}}\right)_{l,4-j}+\left(V_{\tilde{\bm{k}}}\right)_{l+3,4-j}\\
\left(\left(V_{\tilde{\bm{k}}}\right)_{l,4-j}-\left(V_{\tilde{\bm{k}}}\right)_{l+3,4-j}\right)/\iu\\
0
\end{pmatrix},\\
\vec{A}_{j_{1}j_{2}}(\bm{k},\tilde{\bm{q}})&\equiv\frac{1}{\sqrt{3}}\sum_{l=1}^3 e^{-\iu\bm{k}\cdot\bm{d}_{l}}
\left[\left(V_{-\tilde{\bm{q}}}\right)_{l,4-j_{1}}\left(V_{\widetilde{\bm{q}+\bm{k}}}\right)_{l+3,4-j_{2}}
+\left(V_{-\tilde{\bm{q}}}\right)_{l+3,4-j_{1}}\left(V_{\widetilde{\bm{q}+\bm{k}}}\right)_{l,4-j_{2}}\right]\bm{s}_l .
\end{align}
\end{subequations}

The dynamic spin structure factor $\mathcal{S}(\bm{k},\omega)$ shown in this paper is defined as
\begin{equation}
\mathcal{S}(\bm{k},\omega) \equiv \sum_{a=x,y,z} \mathcal{S}^{aa}(\bm{k},\omega),
\end{equation}
and we have approximated the delta functions by Gaussian form with standard deviation $\sigma = \qty{0.015}{meV}$.

{\bf Density matrix renormalization group.}
The dynamic spin structure factor is also calculated by high accuracy density-matrix renormalization group (DMRG), which serves as an unbiased numerical solution for the current problem. 
The cylindrical geometry is used in the DMRG calculation, with a periodic boundary in the $y$ direction and an open boundary in the $x$ direction. We denote it as $L_x\times L_y$ ($L_x\gg L_y$), where $L_x$ and $L_y$ are the number of unit cells in the $x$ and $y$ directions.
We first obtain the ground state by optimizing the matrix product states on the whole cylinder, and then target the dynamical properties (see below) by sweeping the
middle $L_y \times L_y$ unit cells to avoid boundary effect. 
This setup is equivalent to cut the middle $L_y\times L_y$ unit cells and glue them into a torus (with periodic boundary condition along both $x$- and $y$-directions), so that the momentum quantum number can be (approximately) defined along both $x$- and $y$-directions (within $L_y\times L_y$ unit cells in the middle of the cylinder).

To calculate the dynamic spin structure factor, we need to target the following states together with the ground state $|0\rangle$ in the DMRG optimization process:
\begin{center}
\begin{tabular}{lll}
		$|S^\alpha(\bm{k}) \rangle$ & $= S^\alpha(\bm{k}) \,| 0 \rangle$ & 	\\
		$|x^\alpha(\omega+\iu \eta) \rangle$ &$=
	\frac{1}{\omega+\iu \eta-(H-E_{0})} \, {|S^\alpha(\bm{k}) \rangle}$ &
\end{tabular}
\end{center}
where $|x(\omega) \rangle$  is usually called the \textit{correction vector} which can be calculated by the conjugate gradient method~\cite{KuhnerTD1999} or other algorithm~\cite{JeckelmannE2002}.
Using the correction vector, the dynamic spin structure factor can be calculated through:
\begin{equation}
	\mathcal{S}^{\alpha\beta}(\bm{k},\omega) = -\frac{1}{\pi}\text{Im}\langle S^\alpha(\bm{k})|  x^\beta(\omega+\iu \eta) \rangle
\end{equation}
where the smearing energy is set to be $\eta=0.1 J$ in the calculation.
In this work, the calculations were performed on $L_y=6$ cylinders, and we ensure the truncation error is of the order or smaller than $10^{-5}$, by keeping up to $2400$ states in the DMRG process.

\textbf{Acknowledgments}\\
We would like to thank Gang Chen for helpful discussions and  G. Davidson for the great support in setting up and operating the superconducting magnet and the dilution insert throughout the experiment on Pelican.
The authors would also like to acknowledge the neutron beam time awarded by Materials and Life Science Experimental Facility of the Japan Proton Accelerator Research Complex (J-PARC) through Proposal No.~2021B0185 and Australia’s Nuclear Science and Technology Organisation (ANSTO) through proposal No.~P9457. The research was supported by the National Key Research and Development Program of China (Grant No.~2021YFA1400400, 2022YFA1402204), the National Natural Science Foundation of China (Grants No.~12374124, No.~12374146, No.~12104255, No.~12204223), the Open Fund of the China Spallation Neutron Source Songshan Lake Science City (Grant No.~KFKT2023A06). 

\textbf{Author contributions}\\
J.M.S., L.S.W., Z.W. and J.W.M. designed the experiments. W.R.J., L.W. and J.W.M provided single crystals used in this study. J.M.S., M.K., L.S.W and D.H.Y. carried out the neutron scattering experiments. Z.W. carried out the LSW calculations and developed the theoretical explanations. W.Z. performed the DMRG calculations. H.G., N.Z. and T.T.L. carried out the low temperature measurements. All authors discussed the results and contributed to the writing of the manuscript.

\textbf{Competing interests}\\
The authors declare no competing interests.

\newpage

\setcounter{figure}{0}
\setcounter{equation}{0}
\renewcommand{\thefigure}{S\arabic{figure}}
\renewcommand{\thetable}{S\arabic{table}}
\renewcommand\theequation{S\arabic{equation}}

\newpage
\begin{center}
  {\bf ---Supplementary Information---}
\end{center}

\newpage

\begin{figure}[tbp!]
\centering\includegraphics[width=0.65\columnwidth]{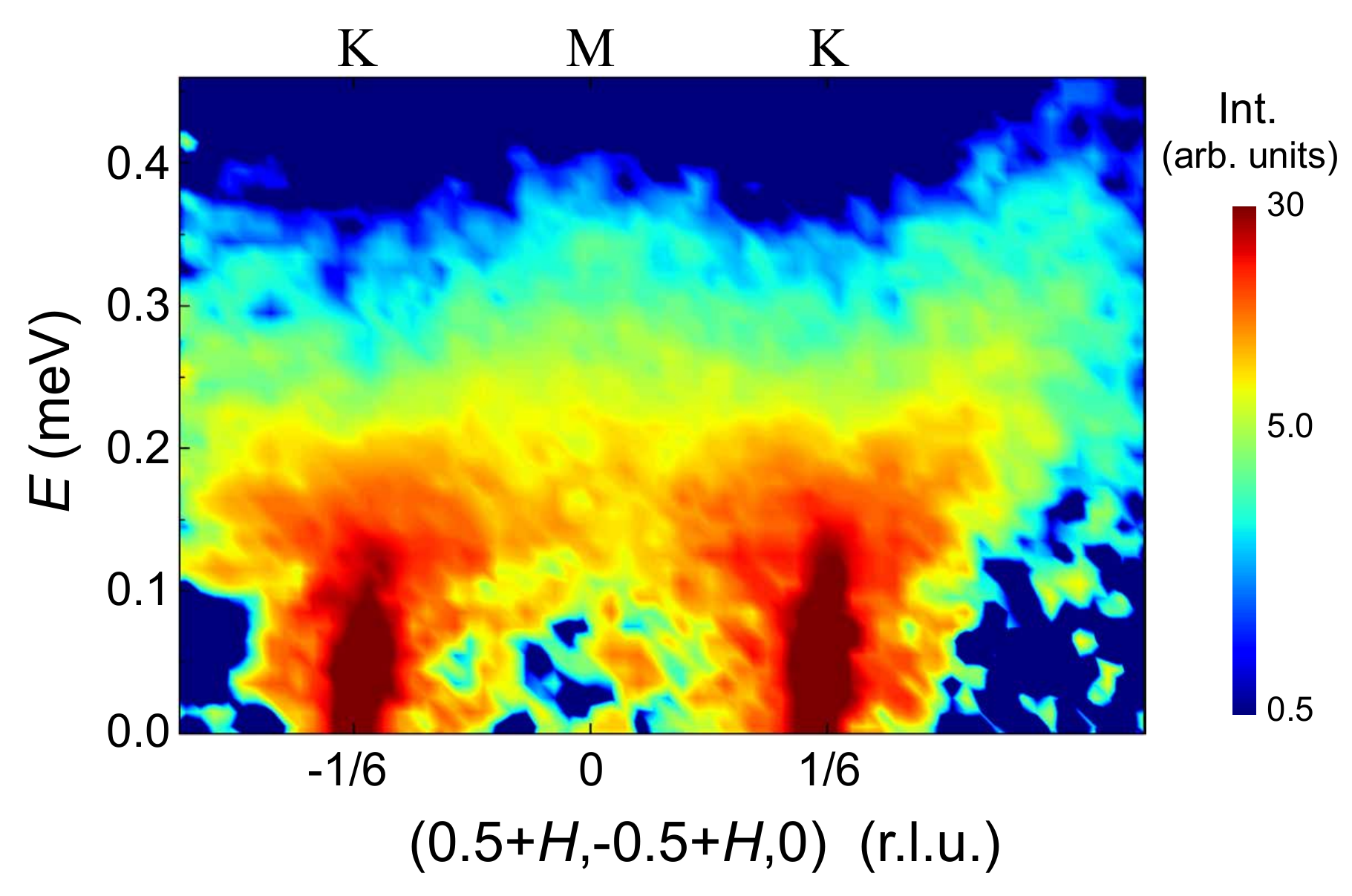} 
\caption{Spin excitation spectrum of \CoP{} measured at $T=\qty{50}{mK}$ and $B=\qty{0}{T}$ along the high symmetry direction using the time-of-flight cold neutron spectrometer, PELICAN, at the OPAL reactor, ANSTO, with a fixed incident energy $E_{\rm i}=\qty{2.63}{meV}$. The energy resolution is about $\qty{0.13}{meV}$.}
\label{KMK_ANSTO}
\end{figure}

Figure~\ref{KMK_ANSTO} is the first measurement showing the spin excitation continuum of \CoP{} along the high symmetry direction at $T=\qty{50}{mK}$ and $B=\qty{0}{T}$ using the time-of-flight cold neutron spectrometer, PELICAN, at the OPAL reactor, ANSTO, with  $E_i=\qty{3.7}{meV}$ and \qty{0.13}{meV} energy resolution. To confirm the continuum nature, we performed further measurements with better energy resolution on AMATERAS at J-PARC.

\begin{figure}[tbp!]
\centering\includegraphics[width=0.99\columnwidth]{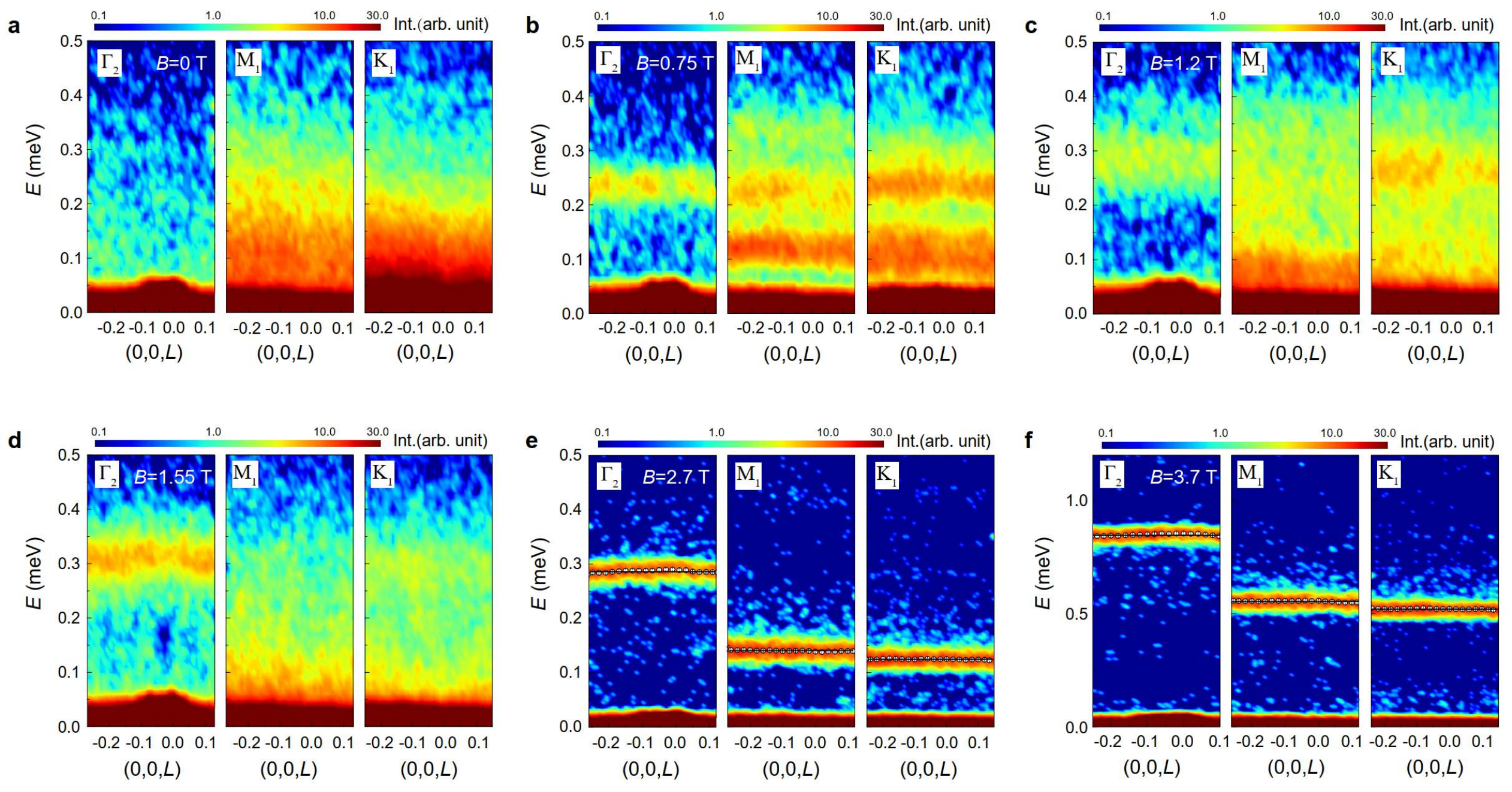} 
\caption{Inelastic neutron excitation spectra of \CoP{} along the (0,0,L) direction at high symmetry points ($\Gamma$, M, and K) measured at $T=\qty{60}{mK}$ and {\bf a} $B=\qty{0}{T}$, {\bf b} $B=\qty{0.75}{T}$, {\bf c} $B=\qty{1.20}{T}$, {\bf d} $B=\qty{1.55}{T}$, {\bf e} $B=\qty{2.70}{T}$, {\bf f} $B=\qty{3.70}{T}$.}
\label{00L_E}
\end{figure}

Figure~\ref{00L_E} presents the inelastic neutron scattering (INS) spectra of \CoP{} along the $[0,0,L]$ direction at high symmetry points ($\Gamma$, M, and K) measured at different magnetic fields and $T=\qty{60}{mK}$ with field applied along the $\bm{c}$-axis. Below the saturation field $B_\text{s}$, the spin excitation spectra exhibit pronounced diffusion, particularly evident in the zero-field where a broad continuum is observed. Conversely, for $B>B_\text{s}$, the spectra manifest as clean and sharply defined spin wave excitations. Overall, the magnetic excitations along the $[0,0,L]$ direction remain nearly flat across all spectra, confirming the quasi-2D nature of the compound.

\begin{figure}[tbp!]
\centering\includegraphics[width=0.75\columnwidth]{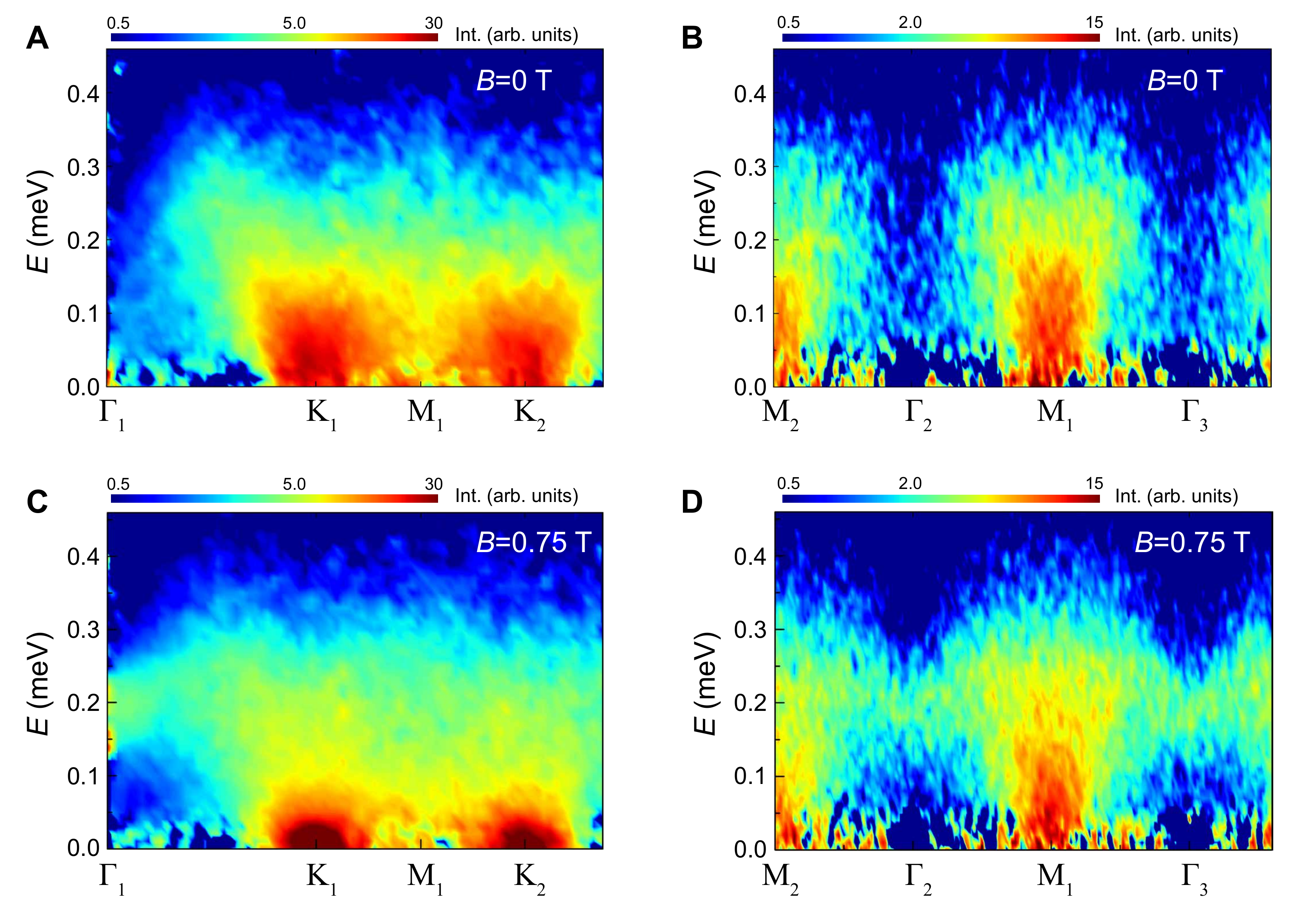} 
\caption{Spin excitation spectrum of \CoP{} measured at $T=\qty{450}{mK}$ along the different high symmetry directions. {\bf a} $\Gamma_1$-K$_1$-M$_1$-K$_2$ at $B=\qty{0}{T}$; {\bf b} M$_2$-$\Gamma_2$-M$_1$-$\Gamma_3$ at $B=\qty{0}{T}$; {\bf c} $\Gamma_1$-K$_1$-M$_1$-K$_2$ at $B=\qty{0.75}{T}$; {\bf d} M$_2$-$\Gamma_2$-M$_1$-$\Gamma_3$ at $B=\qty{0.75}{T}$ using incident neutron energy $E_{\rm i}=\qty{2.63}{meV}$. The intensity is integrated over the window of $L=[-0.2, 0.2]$. All the data was subtracted by a background of $\qty{60}{mK}$-$\qty{3.7}{T}$ data set.}
\label{T450mK_SW3}
\end{figure}

Figure~\ref{T450mK_SW3} displays the spin excitation spectra of \CoP{} along different high-symmetry momentum directions measured at $T=\qty{450}{mK}$ (above $T_\text{N}$) and $B=\qty{0}{T}$, $\qty{0.75}{T}$, respectively. The excitation spectra below and above $T_\text{N}$ are similar at zero field, both manifesting broad continuum excitations, as illustrated in Fig.~4{\bf a}-{\bf b} and Fig.~\ref{T450mK_SW3}{\bf a}-{\bf b}. However, in the UUD phase ($B=\qty{0.75}{T}$), there is a noticeable distinction in the excitation spectra for temperatures below and above $T_\text{N}$. The initially sharp 1-magnon excitation in the UUD phase undergoes rapid diffusion as the temperature crosses $T_\text{N}$, as depicted in Fig.~2{\bf a}-{\bf b} and Fig.~\ref{T450mK_SW3}{\bf c}-{\bf d}. 
This is a natural consequence due to breakdown of long-range magnetic order by thermal fluctuations.

\begin{figure}[tbp!]
\centering\includegraphics[width=0.75\columnwidth]{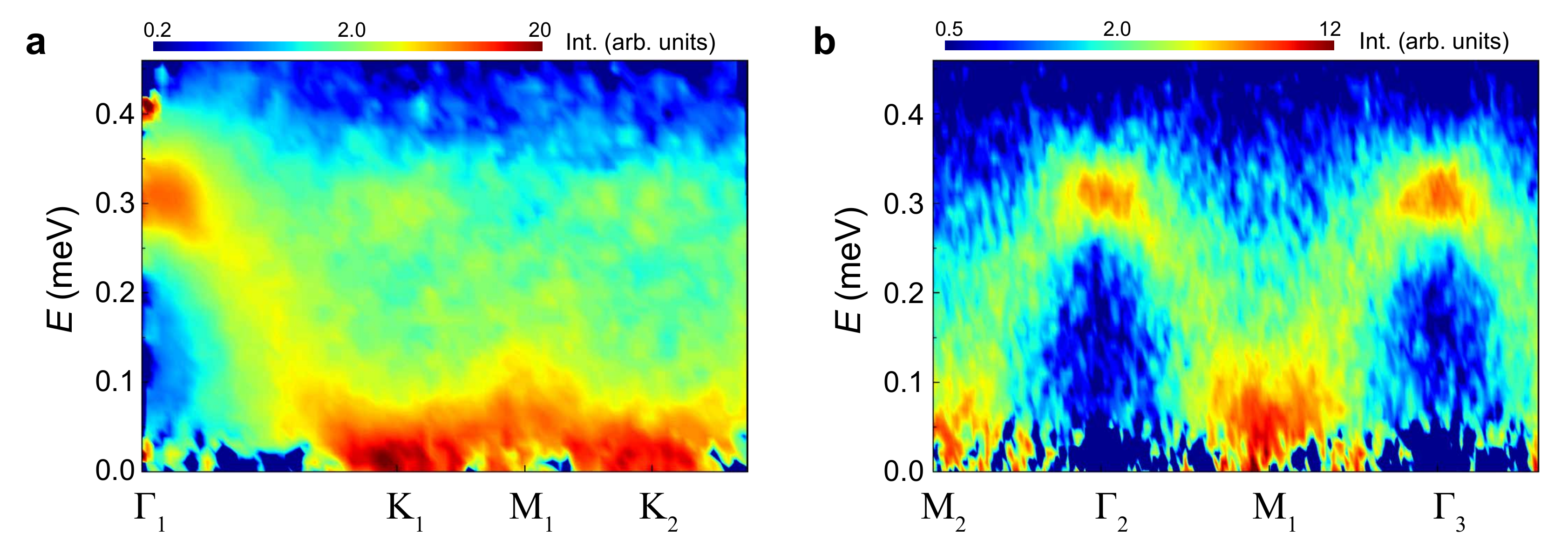} 
\caption{Spin excitation spectra of \CoP{} measured at $T=\qty{60}{mK}$ and $B=\qty{1.55}{T}$ along the different high symmetry directions {\bf a} $\Gamma_1$-K$_1$-M$_1$-K$_2$; {\bf b} M$_2$-$\Gamma_2$-M$_1$-$\Gamma_3$ using incident neutron energy $E_{\rm i}=\qty{2.63}{meV}$. The intensity is integrated over the window of $L=[-0.2, 0.2]$. The data was subtracted by a background of  $\qty{60}{mK}$-$\qty{3.7}{T}$ data set.}
\label{EQ_B1p55T}
\end{figure}

Figure~\ref{EQ_B1p55T} shows the spin excitation spectra of \CoP{} measured at $T=\qty{60}{mK}$ and $B=\qty{1.55}{T}$ along the different high symmetry directions. In addition to the strong 1-magnon spin wave, we observed weak scattering intensity that is broadly distributed across the energy-momentum plane.

\begin{figure*}[t!]
\includegraphics[width=0.99\textwidth]{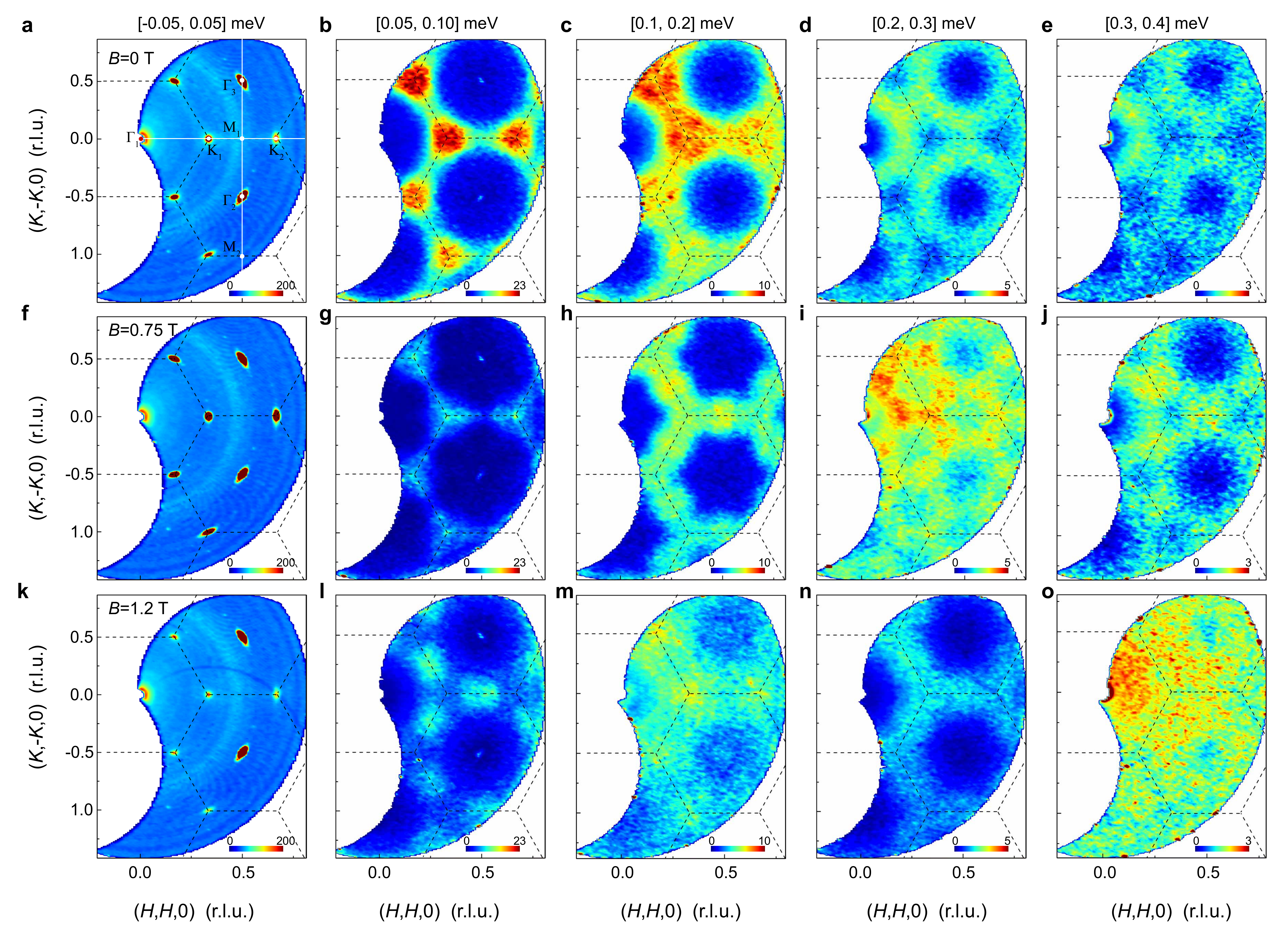}
\caption{Measured momentum dependence of spin excitations in~\CoP{} at different constant energies $E=0,0.075,0.15,0.25,0.35$ meV measured at $T=\qty{60}{mK}$ and {\bf a}-{\bf e} $B=\qty{0}{T}$, {\bf f}-{\bf j} $B=\qty{0.75}{T}$, {\bf k}-{\bf o} $B=\qty{1.2}{T}$. The dashed lines indicate the zone boundaries.}
\label{Eslice}
\end{figure*}


Figure~\ref{Eslice} are the measured momentum dependence of spin excitations of~\CoP{} at different constant energies with $T=\qty{60}{mK}$ and $B=\qty{0}{T}$, $\qty{0.75}{T}$, $\qty{1.2}{T}$, respectively. The evolution of the scattering intensity with increasing energy is clearly observed in these figures. At zero energy, the magnetic Bragg peaks emerge at the K-points for all the magnetic fields. At low energy ($E=\qty{0.075}{meV}$) , strong scattering occurs around the K points for $B=\qty{0}{T}$, while weak intensity is observed around M points for $B=\qty{1.2}{T}$ and the K points for $B=\qty{0.75}{T}$ due to a gap opening at the K points around $\qty{0.05}{meV}$, as illustrated in Fig.~2{\bf a} in the manuscript. As energy increases, the position of strong scattering gradually evolves, and magnetic excitations become increasingly diffusive at high energy due to contribution from the two-magnon continuum.

\end{document}